\documentclass{article}

\usepackage[preprint]{neurips_2023}

\usepackage[utf8]{inputenc} 
\usepackage[T1]{fontenc}    
\usepackage{hyperref}       
\usepackage{url}            
\usepackage{booktabs}       
\usepackage{amsfonts}       
\usepackage{nicefrac}       
\usepackage{microtype}      
\usepackage{xcolor}         
\usepackage{graphicx}
\usepackage{natbib}

\title{EduAgent: Generative Student Agents in Learning}

\author{%
  Songlin Xu$^{1}$~~
  Xinyu Zhang$^{1}$~~
  Lianhui Qin$^{1,2}$\\
  $^1$University of California, San Diego\\
  $^2$Allen Institute for Artificial Intelligence\\
  \texttt{soxu@ucsd.edu} \\
}

\begin{document}

\maketitle



\begin{abstract}

Student simulation in online education is important to address dynamic learning behaviors of students with diverse backgrounds.   
Existing simulation models based on deep learning usually need massive training data, lacking prior knowledge in educational contexts.
Large language models (LLMs) may contain such prior knowledge since they are pre-trained from a large corpus. However, because student behaviors are dynamic and multifaceted with individual differences, directly prompting LLMs is not robust nor accurate enough to capture fine-grained interactions among diverse student personas, learning behaviors, and learning outcomes.
This work tackles this problem by presenting a newly annotated fine-grained large-scale dataset and proposing EduAgent, a novel generative agent framework incorporating cognitive prior knowledge (i.e., theoretical findings revealed in cognitive science) to guide LLMs to first reason correlations among various behaviors and then make simulations.
Our two experiments show that EduAgent could not only mimic and predict learning behaviors of real students but also generate realistic learning behaviors of virtual students without real data.

\end{abstract}

\section{Introduction}
\label{sec: intro}



Online education plays a crucial role not only as a strategic response to a wide variety of disruptions, including natural disasters and public health emergencies  \cite{uscher2018school}, but also as a universally accessible platform to promote inclusivity \cite{mcloughlin2001inclusivity} for students facing challenges in attending traditional in-person classes.
However, online education suffers from several intrinsic limitations that hamper its effectiveness.  In particular, online platforms lack effective mechanisms for the instructor to perceive the students' responses in real time as an ensemble. 
Such perceptions are needed for the instructors to gauge the students' understanding and decide on appropriate lecture adjustments \cite{bond2020mapping,impLearn}.  
``Intelligent tutor'' systems have been proposed to provide feedback to student/teachers, but mostly following hand-crafted rules \cite{Qgen,AutoTutor,zhao2021end}. AI-models can be a powerful alternative, but they either lack real-time responses (e.g., only responding students' final test scores \cite{bassen2020train} or they can only offer ``chatty'' \cite{GPTeach} interactions which are atypical in video lecture-based online education.  

Ideally, the online education should grant instructors a similar or even more granular perception of students' learning behavior, e.g., their engagement, cognitive load, modulated by the course content over time. 
To this end, AI models must overcome two fundamental barriers: (i) Fine-grained learning behavior modeling/prediction; (ii) Acquiring sufficient, labeled data of learning behaviors for model training. 
%
Behavior models can potentially overcome the data scarcity. 
However, existing student simulation models \cite{piech2015deep,beck2000high,hussain2019using,xu2017machine} themselves often need massive training data. 
In this paper, we contribute a new $N = 310$ online education dataset (\textbf{EduAgent310}), consisting of fine-grained, multidimensional records of students' learning behaviors during slide-based lectures. The dataset annotates the students' gaze trajectories, motor control behavior (moving a computer mouse), and 6 different cognitive states. These metrics are timestamped and mapped to the different content blocks on each slide.  Each lecture ends with a comprehensive quiz to evaluate individual students' learning performance. 

The EduAgent310 dataset can provide the ground-truth for modeling learning behaviors. However, a much larger dataset, with a larger student population and more diverse profiles, is needed for training AI-driven pedagogical models.  Unlike common cyber-physical datasets, logging the elusive human cognitive states and behaviors can be extremely time-consuming and costly. 
We thus pose the question: \textit{Can large generative models be used to produce realistic, fine-grained learning behavior data, similar to EduAgent310?} 
To answer this question, we develop a generative agent framework, called EduAgent, which enables us to benchmark the capabilities of state-of-the-art large language models (LLMs) in simulating the fine-grained learning behaviors in response to course content. 

The challenge for the EduAgent framework lies in eliciting the LLM's capability to model sophisticated and dynamic correlation among the students' personas, behaviors, cognitive states, course content, etc. A straightforward prompt is obviously insufficient. 
Advanced LLM-based problem solving techniques, such as Chain of Thought (CoT) \cite{wei2022chain} or few-shot demonstration cannot overcome the challenge either. The learning behavior simulation cannot be easily restructured into step-by-step subtasks which are amenable for CoT.  
On the other hand, few-shot demonstration may either fail to capture the dynamic, multi-faceted student profiles, or overfit to the demo itself.

Our EduAgent framework tackles the challenge by incorporating \textit{\textbf{cognitive priors}}, i.e., classical theories in cognitive science which delineate learning behaviors. 
Specifically, the correlation between personas, course content, and learning behaviors has been well established \cite{karemera2003effects}, but in a piecewise manner. Our EduAgent framework tries to capture the dynamics and the multi-dimensional relation simultaneously, in a modularized architecture which encompasses the different elements in an action space and memory space.   
First, we store different behaviors (such as gaze, cognitive states) in different layers of a memory module. We then prompt the LLM to reason how and why behaviors in each layer are modulated by personas (to capture the individual differences) and course contents (to model the learning process). We also prompt the LLM to reason the correlation between behaviors of different layers, preventing it from overfitting to the few-shot demonstration.
%
%
%
By orchestrating the motor behaviors, persona information, and cognitive states following cognitive prior principles,  
EduAgent can model the learning process in a much finer-grained manner than prior works \cite{chen2023agentverse,jinxin2023cgmi}, thus more accurately predict the learning performance. 

Our experiments on the aforementioned EduAgent310 dataset show that the EduAgent framework can accurately predict a real student's learning behaviors and test results, even with a short history demonstration. 
Furthermore, using the EduAgent framework, 
we generate another dataset (\textbf{EduAgent705}) comprising $N = 705$ virtual students with more diverse personas.  
Our experiments verify that the simulated students exhibit behavioral patterns that are consistent with the real students', and with the cognitive priors, even when no real training data is provided.

In summary, this paper makes three main contributions:


\begin{itemize}
    \item A large-scale and fine-grained newly annotated learning behavior dataset from real students ($N=311$) and virtual students ($N=705$).
    \item An open source generative agent framework {\footnote{Data set and code are available: https://github.com/EduAgent/EduAgent}}, modularized following cognitive priors, to enable realistic simulation of learning behaviors in online education.  
    \item Comprehensive experiments to verify the EduAgent framework and benchmark the capability of SoTA LLMs in modeling fine-grained learning behaviors.
\end{itemize}

Although our current dataset only contains 705 virtual students, the EduAgent framework can be used to generated an unlimited number of virtual students, bearing the cost of accessing the LLM APIs (e.g., \$0.2 or \$0.02 per-student through OpenAI GPT-4 or GPT-3.5). This can potentially overcome the data scarcity bottleneck, enabling the much-anticipated end-to-end human-in-the-loop training of intelligent tutor systems \cite{bhutoria2022personalized}.

\begin{table*}[t]
\caption{Statistics of our dataset compared with existing student learning behavior datasets. \textbf{N}: participant number in the dataset, \textbf{Demo}, \textbf{Gaze}, \textbf{Motor}, \textbf{Cog}, \textbf{Test}, \textbf{Mat} represent whether the dataset contains student personas (demographics or characteristics), gaze behaviors, motor behaviors (such as moving computer mouse in online education or having any gestures in classroom), cognitive states, test question performance, and course/question materials, respectively. \textquotedblleft--\textquotedblright means no explicit information of such data. \textquotedblleft$\times$\textquotedblright represents the lack of such data. \textquotedblleft$\surd$\textquotedblright represents existence of such data. }
\label{t1:dataset}
\vskip 0.15in
\begin{center}
\begin{small}
\begin{sc}
\begin{tabular}{lccccccc}
\toprule
Dataset & N & Demo & Gaze & Motor & Cog & Test & Mat \\
\midrule
\cite{kuzilek2017open} & 32,593 & $\surd$ & $\times$ & $\surd$ & $\times$ & $\surd$ & $\times$ \\
\cite{bui2022online} & 5,327 & $\surd$ & $\times$ & $\times$ & $\times$ & $\surd$ & $\times$ \\
\cite{martin2015student} & 111 & $\surd$ & $\times$ & $\surd$ & $\times$ & $\surd$ & $\times$ \\
\cite{fan2023scb} & -- & $\times$ & $\surd$ & $\surd$ & $\times$ & $\times$ & $\times$ \\
\cite{delgado2021student} & 19 & $\times$ & $\surd$ & $\times$ & $\surd$ & -- & $\times$ \\
\cite{kaur2018prediction} & 78 & $\times$ & $\surd$ & $\times$ & $\surd$ & -- & $\times$ \\
\cite{hasan2021dataset} & 326 & $\surd$ & $\times$ & $\surd$ & $\times$ & $\surd$ & $\times$ \\
\cite{ruiz2022atl} & 54 & $\times$ & $\surd$ & -- & $\surd$ & $\surd$ & $\times$ \\
\cite{mai2022learning} & 400 & -- & $\times$ & $\surd$ & $\times$ & $\surd$ & $\times$ \\
\cite{sun2021student} & -- & $\times$ & $\surd$ & $\surd$ & $\times$ & -- & $\times$ \\
\cite{liu2023xes3g5m} & 18,066 & -- & $\times$ & $\surd$ & $\times$ & $\surd$ & $\surd$ \\
\cite{choi2020ednet} & 1,677,583 & -- & $\times$ & $\surd$ & $\times$ & $\surd$ & $\surd$ \\
\cite{wang2021results} & 118,971 & $\surd$ & $\times$ & $\surd$ & $\times$ & $\surd$ & $\surd$ \\
\cite{stamper20162010} & 1,146 & -- & $\times$ & $\surd$ & $\times$ & $\surd$ & $\surd$ \\
\cite{JunyiOnlineLearningDataset} & 247,606 & $\surd$ & $\times$ & $\times$ & $\times$ & $\surd$ & $\surd$ \\
\cite{statics2011} & 333 & -- & $\times$ & -- & $\times$ & $\surd$ & $\surd$ \\
\cite{feng2009addressing} & 4217 & -- & $\times$ & -- & $\times$ & $\surd$ & $\surd$ \\
EduAgent & 1,015 & $\surd$ & $\surd$ & $\surd$ & $\surd$ & $\surd$ & $\surd$ \\
\bottomrule
\end{tabular}
\end{sc}
\end{small}
\end{center}
\vskip -0.1in
\end{table*}

\section{Related Work}


\subsection{Student Learning Behavior Modelling}
Prior to the maturity of generative agents, substantial research has been devoted to modeling learning behaviors using generic deep learning methods. 
Student learning trace (i.e., records of a student's learning progress) can be modeled using RNN or similar structures \cite{piech2015deep, xiong2016going, chen2018prerequisite, minn2018deep}. Such ``knowledge tracing'' models can be improved by combining exercise contents as well \cite{liu2019ekt}.
Other data-driven approaches have also been widely explored \cite{lee2021prediction, waheed2020predicting} for learning performance prediction.

\subsection{Datasets for Learning Behavior Modeling}

To support student behavior modelling, a variety of datasets have been created.
Table. \ref{t1:dataset} makes a summary for comparison.
Our EduAgent310 and EduAgent705 dataset is unique as it contains fine-grained records of students' cognitive processes in online education. Specifically, it incorporates diverse personas, gaze, motor behaviors, cognitive states, and post-test performance, all synchronized to related course/question materials. 
Mapping between such factors and learning performance has been a long standing problem in cognitive science \cite{resnick2017toward}.
The dataset can enable the development of data-driven models for learning performance prediction, along with real-time feedback/intervention for the students and teachers \cite{pardos2013affective, liu2017investigating}. 




\subsection{LLMs and Agents in Education}

The in-context learning capabilities of LLMs have been harnessed to create emergent agents \cite{lin2023swiftsage} in diverse applications. Examples include content recommendation \cite{zhang2023generative,jin2023lending}, robotic control \cite{ahn2024autort}, web browsing (\cite{yao2022webshop}, \cite{deng2023mind2web}), game player (\cite{gong2023mindagent}), communicative agents (CAMEL \cite{li2023camel}), and so on. 
For human behavior simulation, \cite{park2023generative} explored generative agents for interactive simulacra of human behaviors in a social system \cite{gao2023s3}. \cite{aher2023using} demonstrated the feasibility to replicate human subject studies with LLMs. Finally, \cite{wang2023large} presented an agent framework to simulate user behaviors with memory/actions.


In the education domain, although recent work has explored LLMs to provide feedback to students \cite{kung2023performance,matelsky2023large,cox2023use} and assist teachers \cite{jeon2023large}, 
there is limited research that uses LLM-powered agents to simulate learning behaviors. AgentVerse \cite{chen2023agentverse} simulated classroom 
interactions on their open-sourced agent environments. 
%
CGMI \cite{jinxin2023cgmi} simulated the speech interactions between students and teachers with different personas.  
\cite{xu2023leveraging} leveraged LLMs to simulate cognitive states and learning performance. 
However, no prior work can simulate fine-grained student cognitive states and physiological behaviors. The prompts are usually designed to directly map an input state to learning outcome, abstracting out the correlations among diverse behaviors. 
By contrast, our EduAgent framework models finer-grained physiological and cognitive behaviors, and tackles the long lasting problem of creating realistic cognitive models in an online learning process by integrating cognitive prior knowledge.
EduAgent captures the contextual learning history, students' personas and internal correlations among diverse learning behaviors, which are crucial for cognitive/learning science and AI-driven education.


\begin{figure*}
\centering
\includegraphics[width=0.8\linewidth]{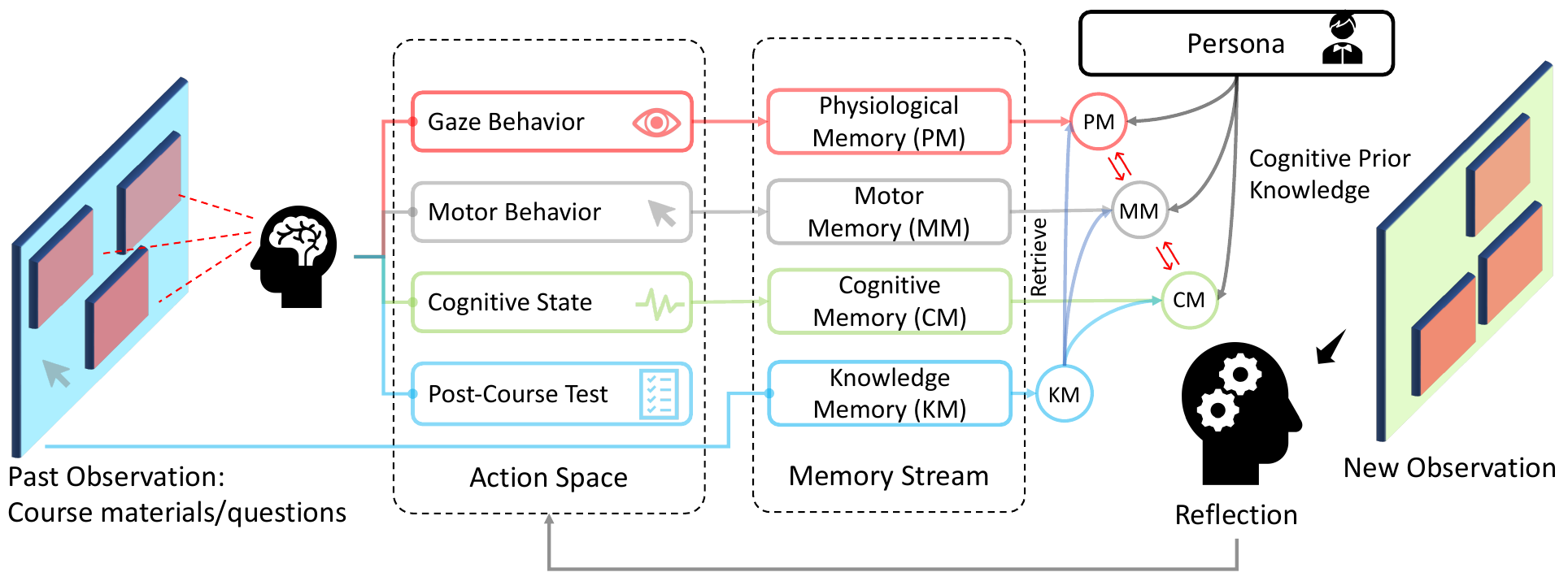}
\caption{Our EduAgent framework. 
}
\label{f1:scenario}
\end{figure*}

\section{Dataset}
\label{sec: dataset}


In this section, we elaborate on the \textbf{EduAgent310} and \textbf{EduAgent705} datasets.

The EduAgent310 is an augmented version from a recent dataset 
\cite{xu2023peer} that contains raw behavior data of $N = 310$ students, where each student watched a slide-based lecture and answered test questions afterwards. The original dataset in \cite{xu2023peer} only contains coarse-grained annotations of student behaviors corresponding to specific post-course questions. 
In contrast, our EduAgent310 adds detailed annotation of gaze/motor behaviors, and cognitive states with precise timestamp. The timestamps are synchronized to the corresponding course materials and post-lecture test questions. 
Each lecture is a 5-min talk about one topic of machine learning and students have diverse educational backgrounds.
Furthermore, for each slide, we annotate the potential Area of Interests (\textbf{AOIs}) in the format of bounding boxes. Each AOI corresponds to a text block, plot, figure illustration, etc. The gaze behavior is a time series of focal points on each slide, measured using the student's webcam and browser-based Webgazer model \cite{xu2023peer}. Motor control behaviors refer to students' mouse moving activities while watching the online course. The time series of gaze and motor data samples are normalized into [0,1], to adapt to different screen size and then mapped into specific AOIs.
%
The cognitive states include workload, curiosity, valid focus, course following, engagement, and confusion. More details about cognitive state measurement are available in Appendix. Each student watched one 5-min video lecture of 30-50 transcripts (sentences) and behaviors are annotated per second, so totally we obtained 100778 labelled samples in EduAgent310 dataset.




To increase the size and diversity of the dataset, we create a new dataset (EduAgent705) composed of $N = 705$ virtual students.   
The virtual students are simulated by our EduAgent framework and verified through experiments (Section \ref{sec: exp2}). Before elaborating on the framework design, we first introduce the dataset itself. 
Inspired by \cite{seidel2021student,nakayama2021impact} that shows the effect of the students' personas on learning performance, we consider the following \textit{personas} when simulating virtual students:
learning attitude, exam performance, focus, curiosity, interest in course, prior course knowledge, compliance in course, smartness, and family. Each characteristic has one positive item and one negative item, listed in Appendix Table. \ref{appendix table:demo virtual student}.
We further include demographic information such as age, major and education levels, listed in Appendix Table. \ref{appendix table:demo virtual student}. Totally, there are $4\times3\times6\times4\times2^9=147456$ possible combinations of personas.  Before running the EduAgent simulation, we instantiate the framework using one randomly selected persona.  Our current simulation has created 705 virtual students, each going through one lecture session and one post-lecture test. But the framework itself is capable of generating unlimited amount of samples. 


\section{EduAgent Framework}
\label{sec: framework}

\begin{table*}[t]
\caption{Micro benchmark to show results in the first experiment. $\surd$: with cognitive prior knowledge, $\times$: no cognitive prior knowledge. \textbf{Cop.} means components used in the ablation study. \textbf{All}: all components are used, \textbf{$\times$M}: motor behaviors are removed in the memory, \textbf{$\times$P}: gaze behaviors are removed in the memory, \textbf{$\times$C}: cognitive states are removed in the memory, and \textbf{$\times$D}: the whole few-shot memory as example demonstrations are removed. For metrics, \textbf{Ga.}: gaze AOI distance, \textbf{Mo.}: motor AOI distance, \textbf{Wo.}: workload MAE, \textbf{Cu.}: curiosity MAE, \textbf{Foc.}: valid focus MAE, \textbf{Fol.}: course following MAE, \textbf{Eng.}: engagement MAE, \textbf{Co.}: confusion MAE, \textbf{CH.}: choice similarity, \textbf{Ac.}: accuracy similarity. \textcolor{blue}{Blue} color means the configuration in the current row could achieve \textcolor{blue}{\textbf{better}} simulation performance compared with the first row (GPT3.5 with cognitive prior knowledge with all components in the framework) in the specific metric while \textcolor{red}{red} color means the configuration in the current row results in \textcolor{red}{\textbf{worse}} performance compared with the first row.
}
\label{t3:ablation of components}
\vskip 0.15in
\begin{center}
\begin{small}
\begin{sc}
\begin{tabular}{lcccccccccccc}
\toprule
Model & Pri. & Cop. & Ga. & Mo. & Wo. & Cu. & Foc. & Fol. & Eng. & Co. & Ch. & Ac. \\
\midrule
GPT3.5 & $\surd$ & All & \textcolor{black}{0.35} & 0.34 & \textcolor{black}{0.17} & \textcolor{black}{0.23} & \textcolor{black}{0.25} & \textcolor{black}{0.35} & \textcolor{black}{0.11} & \textcolor{black}{0.07} & \textcolor{black}{0.61} & \textcolor{black}{0.66}\\
GPT3.5 & $\times$ & All & \textcolor{red}{0.35} & \textcolor{blue}{0.34} & \textcolor{red}{0.25} & \textcolor{red}{0.29} & \textcolor{red}{0.27} & \textcolor{red}{0.39} & \textcolor{red}{0.18} & \textcolor{red}{0.11} & \textcolor{red}{0.60} & \textcolor{red}{0.65} \\ 
\hline
GPT3.5 & $\surd$ & $\times$M & \textcolor{red}{0.35} & \textcolor{red}{0.35} & \textcolor{red}{0.17} & \textcolor{blue}{0.22} & \textcolor{red}{0.25} & \textcolor{blue}{0.34} & \textcolor{blue}{0.10} & \textcolor{blue}{0.06} & \textcolor{red}{0.60} & \textcolor{red}{0.65}\\
GPT3.5 & $\surd$ & $\times$P & \textcolor{red}{0.37} & \textcolor{red}{0.35} & \textcolor{red}{0.18} & \textcolor{red}{0.23} & \textcolor{red}{0.25} & \textcolor{blue}{0.34} & \textcolor{red}{0.12} & \textcolor{blue}{0.07} & \textcolor{red}{0.60} & \textcolor{red}{0.65}\\
GPT3.5 & $\surd$ & $\times$C & \textcolor{red}{0.35} & \textcolor{red}{0.34} & \textcolor{red}{0.17} & \textcolor{red}{0.48} & \textcolor{red}{0.27} & \textcolor{red}{0.56} & \textcolor{red}{0.26} & \textcolor{red}{0.19} & \textcolor{red}{0.60} & \textcolor{red}{0.65}\\
GPT3.5 & $\surd$ & $\times$D & \textcolor{red}{0.36} & \textcolor{red}{0.34} & \textcolor{red}{0.17} & \textcolor{red}{0.50} & \textcolor{red}{0.27} & \textcolor{red}{0.55} & \textcolor{red}{0.28} & \textcolor{red}{0.19} & \textcolor{red}{0.56} & \textcolor{red}{0.63}\\
\hline
GPT4 & $\surd$ & All & \textcolor{blue}{0.35} & \textcolor{blue}{0.32} & \textcolor{red}{0.20} & \textcolor{red}{0.40} & \textcolor{red}{0.26} & \textcolor{red}{0.55} & \textcolor{red}{0.12} & \textcolor{red}{0.15} & \textcolor{blue}{0.66} & \textcolor{blue}{0.68}\\
Gemini & $\surd$ & All & \textcolor{red}{0.37} & \textcolor{blue}{0.34} & \textcolor{red}{0.21} & \textcolor{red}{0.26} & \textcolor{blue}{0.23} & \textcolor{red}{0.43} & \textcolor{blue}{0.03} & \textcolor{blue}{0.02} & \textcolor{red}{0.57} & \textcolor{red}{0.60}\\
LLAMA2 & $\surd$ & All & \textcolor{red}{0.36} & \textcolor{blue}{0.32} & \textcolor{red}{0.28} & \textcolor{red}{0.32} & \textcolor{red}{0.27} & \textcolor{blue}{0.34} & \textcolor{blue}{0.04} & \textcolor{red}{0.14} & \textcolor{red}{0.39} & \textcolor{red}{0.52}\\
\bottomrule
\end{tabular}
\end{sc}
\end{small}
\end{center}
\vskip -0.1in
\end{table*}

\subsection{Our Approach}
The simulation pipeline is depicted in Fig. \ref{f1:scenario}. Specifically,
we first instantiate each student agent with one persona profile as mentioned before. Then we simulate the agent's learning process from the first to the last slide of the lecture in order. Each slide is one simulation step, where the agent receives the transcripts within this slide and outputs a trajectory of simulated learning behaviors (actions) for each transcript (each transcript is one sentence). The actions include the student's gaze behaviors, motor control behaviors to move computer mouse, and cognitive states (workload, curiosity, valid focus, course following, engagement, confusion), as well as answers of corresponding test questions related to the specific slide. 
Before making actions, agents first reflect the correlation among personas, past actions, and past course materials from the memory (demonstrations), guided by the integrated cognitive priors (depicted below). The demonstrations give the agent its past gaze/motor behaviors, cognitive states and past course materials in the time series format so that it can reason how different behaviors affect each other (reflection outcomes). Note that we only use past behaviors of the most recent past slide (not all past slides) for few-shot demonstration.
After that, agents apply the reflection outcomes on new input course materials or test questions to make actions.

A key design principle of EduAgent is to incorporate cognitive priors \cite{bourgin2019cognitive}, which helps guide the LLMs to \textbf{first reason correlations} among learning behaviors and \textbf{then make simulations}. 
At a high level, we glean theoretical findings of student learning behaviors in cognitive science (e.g., correlations among cognitive states and learning performance), and embed such prior knowledge in the prompts, so that the LLM can stay grounded and get clear guidance regarding where to start reasoning.
Instead of giving an explicit statement of the cognitive priors, we allow the LLM to reason by itself, regarding how and why behaviors in each memory layer are modulated by personas (to capture individual differences) and course contents (to model learning process). We also prompt the LLM to reason how different behaviors affect each other, preventing it from directly copying few-shot history demonstration and mitigating the overfitting problem. 
\textbf{Gaze/Motor Behavior Simulation}: 
It is difficult for the LLM to directly interpret raw gaze/motor sensor data, because they are usually noisy and massive, exceeding token limitation and lacking contextual information.
%
Theoretical studies of online education \cite{massaro2012art} have established the correlation between gaze/motor behavior and the lecture content. Such correlation effects are multifaceted, vary over time, and do not admit any closed-form model. However, Mayer et al. advocate that effective online learning occurs when a student selects relevant elements, organizes the elements to form coherent mental representations, and integrates the new and existing representations \cite{MayerBook}. This process necessarily involves the interaction between gaze/motor following behaviors and the semantics within the course content, where LLM excels at. 
We thus propose to map gaze/motor coordinates into specific AOIs on slides, so that LLMs can correlate the semantic information to sensory behaviors. Instead of asking the agent to output the raw gaze/motor data, we prompt it to output the gaze/motor AOI ID on each slide. Gaze/motor changes across the AOIs in turn serve as the action for our gaze/motor simulation.
%
%
During a reflection, 
we prompt the agent to first reason, based on its memory, regarding how these factors modulate gaze/motor behaviors. The agent then leverages the reasoning outcomes to perform the gaze/motor simulation 
according to new course materials.











\textbf{Cognitive States Simulation}: 
For each specific transcript, the agent generates a numeric value ranging from 0 to 1 to indicate the level of each cognitive state factor. 
\cite{furnham2003personality} revealed that student cognitive states are not only modulated by course materials themselves, but also affected by students' own personas. For example, students who do not have strong academic background may have higher workload in course. Moreover, \cite{d2012gaze} showed student gaze/motor behaviors can be indicators of cognitive states during learning.
%
These cognitive priors inspire us to prompt the agent to first reason how its persona and past course contents modulate its past cognitive states and how past gaze/motor behaviors 
can indicate cognitive states from demonstrations in memory. 
The agent then applies such reasoning outcomes on current course materials and simulated gaze/motor behaviors to estimate the modulated cognitive states for output.

\textbf{Learning Performance Simulation}: For each post-lecture test question, the agent makes one selection from four choices. The goal is to mimic the question answering of each individual student instead of directly choosing the correct answer. 
Using two longitudinal university samples, \cite{chamorro2003personality} revealed that personality can serve as important predictors of student academic performance. Moreover, \cite{zhu2023integrating} showed that gaze/motor behaviors are strongly correlated with student activities in e-learning to infer learning performance. \cite{lei2018relationships} further revealed that student engagement is apparently correlated with student learning performance. 
These cognitive priors inspire us to prompt the agent to reason how its persona and simulated cognitive states and gaze/motor behaviors could affect its question answer correctness according to course materials.
%
If the agent reasons that the student is likely to choose the correct answer, it finds the correct answer based on transcripts. Otherwise, it should estimate the most likely but incorrect choice according to gaze/motor trajectories across the transcripts.

\section{Experiment 1: Personalized Student Behavior Prediction}
\label{sec: evaluation real}
\subsection{Task Settings and Metrics}
We first evaluate EduAgent's ability to predict future student learning behaviors and outcomes, using students' past learning behavior as few-shot demonstration. In our experiment, we simulate 310 student agents, corresponding to the $N = 310$ dataset (\textbf{EduAgent310}). Each agent experiences 30-50 transcripts and takes actions per transcript.
%
As depicted in Section. \ref{sec: framework}, we simulate students per slide and give corresponding course materials and post questions as input for student agents. 
After each slide, we log the real students' behaviors from previous slides into the agents' memory, to personalize their responses for future slide simulation. 

To compare behaviors between agents and corresponding real students (ground truth), we use normalized \textbf{AOI distance} on screen for gaze/motor behaviors, \textbf{Mean Absolute Error} (MAE) for six cognitive states (normalized to [0,1]), \textbf{choice similarity} (whether both choices are similar) and \textbf{accuracy similarity} (whether both accuracy is similar) for question answering simulation. More details about metric design are depicted in Appendix.


\begin{figure*}
\centering
\includegraphics[width=0.8\linewidth]{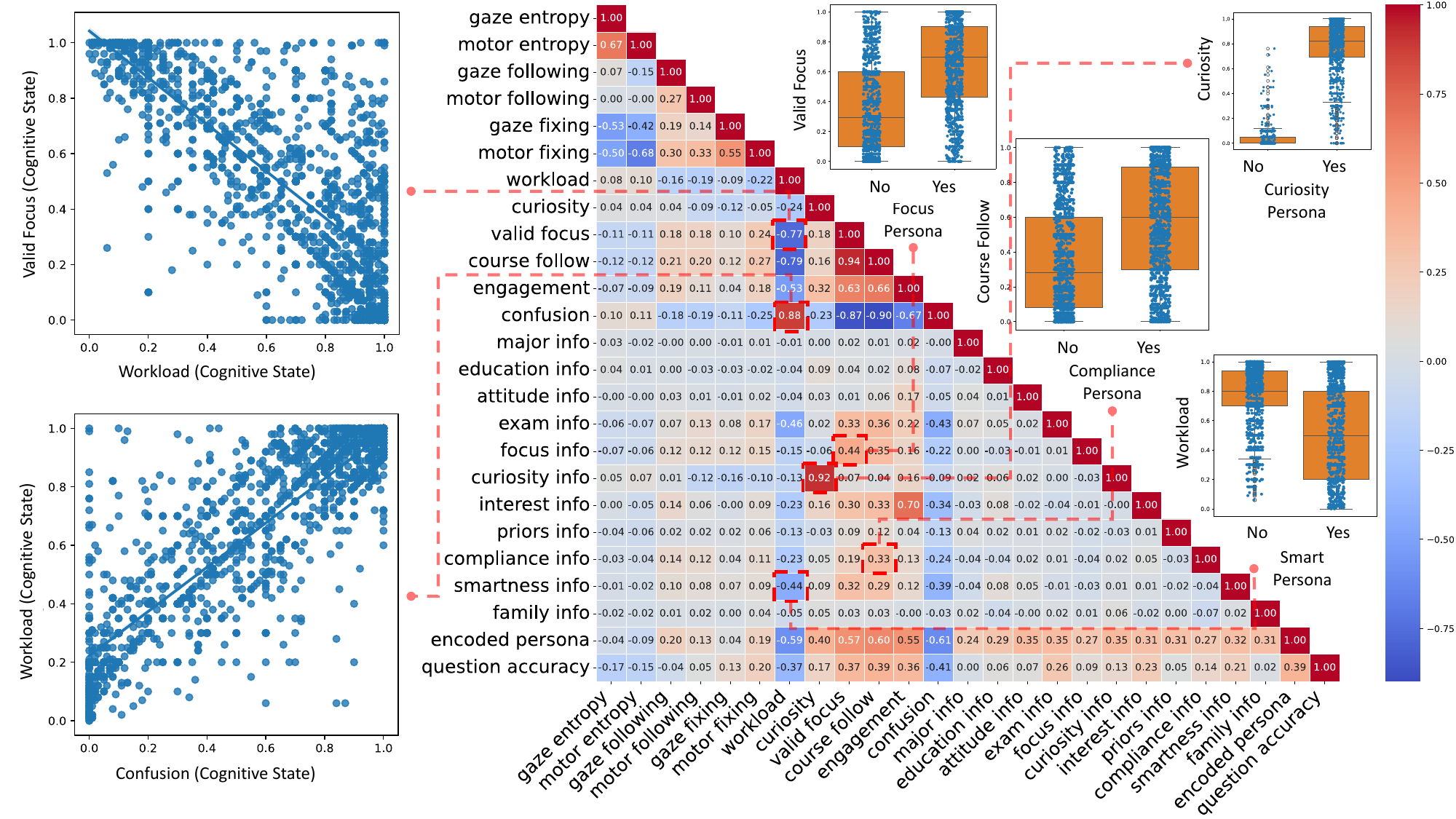}
\caption{Correlation matrix heatmap and example relationships of generated behaviors in the second experiment.
}
\label{f5:correlation heatmap}
\end{figure*}

\subsection{Results and Analysis}
For all language model generations, we set temperature to be 0 for more deterministic results.




\textbf{Importance of Cognitive Prior Knowledge}: We first verify whether cognitive prior knowledge can improve simulation performance using OpenAI GPT-3.5 model\cite{OpenAI}. As depicted in Section. \ref{sec: intro}, it is hard to directly apply recent advances of prompting techniques, such as Chain of Thought (CoT) \cite{wei2022chain} into our problem. Therefore, we use standard prompt to serve as baseline that directly asks LLMs
to give the output 
(actions) based on course contents, questions and memory.
The results in Table. \ref{t3:ablation of components} show that integrating cognitive priors improves simulation performance including gaze behaviors, cognitive states and question answering. For motor behaviors, standard prompt is slightly better (AOI Distance = 0.336) compared with integrating cognitive priors (AOI Distance = 0.340). One potential reason is that, unlike gaze behaviors that indicate explicit student focus, mouse moving behaviors show weak correlations with other behaviors. Therefore integrating cognitive priors may not significantly enhance them.
However, cognitive priors significantly improve simulation performance on cognitive states.
These results indicate that, by incorporating cognitive prior knowledge to give clear guidelines for agents to reason from memory, agents can better capture potential correlations among diverse behaviors, and therefore further improve simulation. This also provides insights for future agent framework design (not limited to student agents but also other generative agents) that incorporating prior knowledge about correlations among actions and observations may boost agents' performance.

\textbf{Importance of Different Components}: We also run ablation studies to compare the importance of different components in our EduAgent framework. Specifically, we remove specific behavior data from memory and compare the performance difference. 
As depicted in Table. \ref{t3:ablation of components}, we find that all behavior simulation performance drops when we remove past cognitive states from memory, indicating that cognitive states are highly correlated with other behaviors and therefore play an important role in our framework to provide contextual information for simulation in all behaviors. We also find that gaze/motor and question answering simulation performance drops more heavily when removing past gaze behaviors from memory compared with removing cognitive states. These results indicate that student gaze/motor and learning performance are more correlated with past gaze behaviors compared with past cognitive states, demonstrating the \textbf{importance of incorporating physiological behaviors} for student simulation, which is also a distinguishing feature of \textbf{our datasets} compared with other existing datasets. We also find that removing past motor behaviors from memory results in slightly better performance in four cognitive states simulation. This echoes the previous explanation comparing with standard prompts showing that mouse moving behaviors do not have that strong correlations with other behaviors. We also explore the effect when removing the whole few-shot memory as example demonstration and we find that the simulation performance drops significantly for all behaviors, indicating that a student's historical data (the few-shot example demonstration) plays an important role in personalizing the student agent for behavior simulation. 




\textbf{Impact of Foundation Models}: We also compare performance with different foundation models. 
As depicted in Table. \ref{t3:ablation of components}, with Gemini Pro \cite{Gemini} and Llama 2 70B \cite{llama2-70b}, we find slight improvement in motor and cognitive states simulation but obvious performance drop in gaze and question answering simulation compared with GPT-3.5, indicating the stronger ability of GPT-3.5 to capture correlation between student learning behaviors and learning performance. Moreover, the obvious improvement in gaze/motor and question answering simulation with GPT-4 \cite{OpenAI} suggests its stronger ability than GPT-3.5 to capture such correlation. 

\section{Experiment 2: Virtual Generative Student Simulation}
\label{sec: exp2}

\subsection{Task Settings and Metrics}
Our second experiment tests whether EduAgent exhibits reasonable learning behavioral patterns without real demonstration data. The experiment is conducted using the EduAgent705 dataset, based on the optimal configuration we have identified in the first experiment (i.e., GPT-4 with cognitive priors with all components). Since there is no real student data, the agents directly use their own generated past behaviors as memory. 
%

For virtual student datasets, we do not have ground truth for comparison. 
Therefore, we use Pearson coefficients to measure 
correlation among personas (demographics and characteristics) and behaviors, By doing so, we can verify whether such simulated behaviors echo related findings in cognitive science. 
To facilitate the Pearson coefficient calculation, we encode student specific personas into numeric values, and use entropy/following/fixing to represent different aspects of gaze/motor patterns (details in Appendix).
\subsection{Results and Analysis}

Detailed results of the 705 simulated agents are depicted in Fig. \ref{f5:correlation heatmap}. 
Overall, most behaviors are consistent with well established principles in cognitive and learning science. We elabrate on a few representatives below (Pearson coefficients denoted by $r$).


\textbf{Persona v.s. Gaze/Motor}: 
We find that student agents with high GPA in exam have better gaze/motor following (gaze: $r=0.07$, motor: $r=0.13$) and fixing behaviors (gaze: $r=0.08$, motor: $r=0.17$) but low entropy (gaze: $r=-0.06$, motor: $r=-0.07$). Similar correlation can be found between focused/compliance persona and gaze/motor behaviors. By contrast, agents who have curious persona have larger gaze/motor entropy (gaze: $r=0.05$, motor: $r=0.07$) but smaller motor following ($r=-0.12$) and gaze/motor fixing (gaze: $r=-0.16$, motor: $r=-0.10$). 
%
Such results echo \cite{kosel2021identifying} about the correlation between gaze and student characteristics. Specifically, more compliant/attentive students have lower gaze entropy and better following/fixing since they are more focused in class. The results also verifies the effectiveness of the EduAgent framework in incorporating cognitive priors as well as knowledge of the students' personas. 

\textbf{Persona v.s. Cognitive States}: We find that agents with high learning curiosity personas are also highly curious in cognitive states($r=0.92$). Smart personas correspond to low course workload ($r=-0.44$) and attentive personas lead to better focus ($r=0.44$). Finally, we find a positive correlation between agents GPA information and their valid focus ($r=0.33$), course following ($r=0.36$), and engagement ($r=0.22$), but negative correlation between GPA and workload ($r=-0.46$), confusion($r=-0.43$). 
Such results verify the effectiveness of the EduAgent simulation, which echoes prior research \cite{lau2002cognitive} in correlating learners' characteristics and cognitive states. 

\textbf{Persona v.s. Question Answering}: 
We find correlation between answer accuracy and persona ($r=0.39$), in education level ($r=0.06$), attitude ($r=0.07$), GPA ($r=0.26$), focus ($r=0.09$), curiosity ($r=0.13$), interest ($r=0.23$), prior knowledge ($r=0.05$), compliance ($r=0.14$), and smartness ($r=0.21$).
%
The results echo \cite{karemera2003effects} in correlation between characteristics and academic performance and reflect cognitive prior knowledge design to consider personas impact on question answering. 

\textbf{Cognitive States v.s. Gaze/Motor}: We find engagement is positively correlated with following (gaze: $r=0.19$, motor: $r=0.11$) and fixing behaviors (gaze: $r=0.04$, motor: $r=0.18$) but negatively correlated with entropy (gaze: $r=-0.07$, motor: $r=-0.09$). Similar relationships are also found between valid focus/course following and gaze/motor behaviors. By contrast, we can also find opposite relationships between workload/curiosity and gaze/motor behaviors.
These results are aligned with theories in cognitive science \cite{d2012gaze},  and are \textbf{promising because} gaze/motor patterns are calculated from \textbf{actions of gaze/motor AOI} by comparing with course pace AOI while curiosity/workload/etc are obtained from \textbf{actions of cognitive states}. Such consistency suggests that our EduAgent successfully establishes the mapping between gaze/motor behaviors and cognitive states.
%

\textbf{Gaze v.s. Motor}: We find consistency between gaze and motor behaviors in most cases. Specifically, we find positive correlation between gaze and motor in entropy ($r=0.67$),  following ($r=0.27$), and fixing ($r=0.55$) patterns. Moreover, they do not fully overlap since gaze behaviors represent explicit watching focus while motor behaviors represent implicit focus by mouse moving \cite{guo2010towards}.
The results echo \cite{zhang2020data,zhu2023integrating} about synchronization between students gaze and motor behaviors because mouse movement driven by cognitive states like curiosity or engagement could be indicated from gaze patterns \cite{kwok2018understanding}. 

\textbf{Gaze/Motor v.s. Question Answering}: We find that answer accuracy is negatively correlated with gaze ($r=-0.17$) and motor entropy ($r=-0.15$) but positively correlated with gaze ($r=0.13$) and motor fixing ($r=0.20$). 
The results reflect our cognitive prior knowledge design to consider such correlation and echo \cite{zhu2023integrating} since gaze/motor behaviors indicate cognitive states and therefore reveal learning success \cite{d2012gaze}. However, we also find weak yet opposite effect of gaze ($r=-0.04$) and motor following ($r=0.05$). This reflects the relationship between gaze and motor behaviors discussed above, i.e. gaze and motor behaviors do not exactly overlap although they have closed correlation \cite{guo2010towards}.

\textbf{Cognitive States v.s. Question Answering}: We find that answer accuracy is negatively correlated with workload ($r=-0.37$) and confusion ($r=-0.41$), and positively correlated with curiosity ($r=0.17$), valid focus ($r=0.37$), course following ($r=0.39$), engagement ($r=0.36$).
Such results echo \cite{lei2018relationships} about the correlations between cognitive states and academic performance and reflect our cognitive prior knowledge design to consider interactions between cognitive states and question answering. These results are \textbf{promising because} question accuracy is calculated from \textbf{actions of agent choices} by comparing with correct answers while curiosity/workload/etc are obtained from \textbf{actions of cognitive states}. Therefore, such consistency suggest that agents successfully map the correlation between cognitive states and learning success.

\section{Limitations and Discussion}
Student behavioral simulation is challenging as human behaviors are dynamic and come with noise in nature\cite{cziko1989unpredictability}. 
For example, it is easy for LLMs to answer questions correctly according to course materials. But it is hard to predict the same wrong choice of students. 
Moreover, except for precisely predicting student behaviors, we suggest that generating realistic learning behaviors like the second experiment is also one important research problem and may have broader impacts such as providing realistic course feedback to improve teaching strategy/course quality and facilitating hypotheses exploration in educational research.
However, unlike directly asking LLMs to mimic specific personas or behaviors in individual cases, our problem mixes all student behaviors, personas and course contents together. Therefore, it is challenging for LLMs to cover all potential correlations while taking such massive information as input. Hence, there are also some inconsistent correlations as depicted in the second experiment. Future work could explore how to further improve the simulation performance.



%


\section{Conclusion}

We propose a novel generative agent framework (EduAgent) to simulate fine-grained and comprehensive student learning behaviors in online education. Two datasets are contributed to facilitate generative student agent research. Our experiments show that LLMs could not only predict student learning behaviors according to personalized history, but also generate realistic learning behavioral patterns without real data. These results suggest a promising new line of research in student learning behavioral modelling and generative student agents. We believe that our work could serve as important groundwork and provide new insights in both student simulation and educational research.

\bibliography{0_reference}

\begin{thebibliography}{86}
\providecommand{\natexlab}[1]{#1}
\providecommand{\url}[1]{\texttt{#1}}
\expandafter\ifx\csname urlstyle\endcsname\relax
  \providecommand{\doi}[1]{doi: #1}\else
  \providecommand{\doi}{doi: \begingroup \urlstyle{rm}\Url}\fi

\bibitem[Gem()]{Gemini}
Gemini pro.
\newblock \url{https://ai.google.dev/tutorials/python_quickstart}.
\newblock Accessed: January 24, 2024.

\bibitem[Ope()]{OpenAI}
Openai.
\newblock \url{https://platform.openai.com/docs/models}.
\newblock Accessed: January 24, 2024.

\bibitem[lla()]{llama2-70b}
Llama 2 70b.
\newblock \url{https://huggingface.co/meta-llama/Llama-2-70b-chat-hf}.
\newblock Accessed: January 24, 2024.

\bibitem[sta()]{statics2011}
Statics2011 dataset.
\newblock \url{https://pslcdatashop.web.cmu.edu/DatasetInfo?datasetId=507}.
\newblock Accessed: January 24, 2024.

\bibitem[Aher et~al.(2023)Aher, Arriaga, and Kalai]{aher2023using}
Gati~V Aher, Rosa~I Arriaga, and Adam~Tauman Kalai.
\newblock Using large language models to simulate multiple humans and replicate human subject studies.
\newblock In \emph{International Conference on Machine Learning}, pages 337--371. PMLR, 2023.

\bibitem[Ahn et~al.(2024)Ahn, Dwibedi, Finn, Arenas, Gopalakrishnan, Hausman, Ichter, Irpan, Joshi, Julian, Kirmani, Leal, Lee, Levine, Lu, Leal, Maddineni, Rao, Sadigh, Sanketi, Sermanet, Vuong, Welker, Xia, Xiao, Xu, Xu, and Xu]{ahn2024autort}
Michael Ahn, Debidatta Dwibedi, Chelsea Finn, Montse~Gonzalez Arenas, Keerthana Gopalakrishnan, Karol Hausman, Brian Ichter, Alex Irpan, Nikhil Joshi, Ryan Julian, Sean Kirmani, Isabel Leal, Edward Lee, Sergey Levine, Yao Lu, Isabel Leal, Sharath Maddineni, Kanishka Rao, Dorsa Sadigh, Pannag Sanketi, Pierre Sermanet, Quan Vuong, Stefan Welker, Fei Xia, Ted Xiao, Peng Xu, Steve Xu, and Zhuo Xu.
\newblock Autort: Embodied foundation models for large scale orchestration of robotic agents, 2024.

\bibitem[Asteriadis et~al.(2009)Asteriadis, Tzouveli, Karpouzis, and Kollias]{asteriadis2009estimation}
Stylianos Asteriadis, Paraskevi Tzouveli, Kostas Karpouzis, and Stefanos Kollias.
\newblock Estimation of behavioral user state based on eye gaze and head pose—application in an e-learning environment.
\newblock \emph{Multimedia Tools and Applications}, 41:\penalty0 469--493, 2009.

\bibitem[Bassen et~al.(2020)Bassen, Balaji, Schaarschmidt, Thille, Painter, Zimmaro, Games, Fast, and Mitchell]{bassen2020train}
Jonathan Bassen, Bharathan Balaji, Michael Schaarschmidt, Candace Thille, Jay Painter, Dawn Zimmaro, Alex Games, Ethan Fast, and John~C Mitchell.
\newblock How to train your learners: Reinforcement learning for the scheduling of online learning activities.
\newblock 2020.

\bibitem[Beck and Woolf(2000)]{beck2000high}
Joseph~E Beck and Beverly~Park Woolf.
\newblock High-level student modeling with machine learning.
\newblock In \emph{International Conference on Intelligent Tutoring Systems}, pages 584--593. Springer, 2000.

\bibitem[Bhutoria(2022)]{bhutoria2022personalized}
Aditi Bhutoria.
\newblock Personalized education and artificial intelligence in the united states, china, and india: A systematic review using a human-in-the-loop model.
\newblock \emph{Computers and Education: Artificial Intelligence}, 3:\penalty0 100068, 2022.

\bibitem[Bond et~al.(2020)Bond, Buntins, Bedenlier, Zawacki-Richter, and Kerres]{bond2020mapping}
Melissa Bond, Katja Buntins, Svenja Bedenlier, Olaf Zawacki-Richter, and Michael Kerres.
\newblock Mapping research in student engagement and educational technology in higher education: A systematic evidence map.
\newblock \emph{International journal of educational technology in higher education}, 17\penalty0 (1), 2020.

\bibitem[Bourgin et~al.(2019)Bourgin, Peterson, Reichman, Russell, and Griffiths]{bourgin2019cognitive}
David~D Bourgin, Joshua~C Peterson, Daniel Reichman, Stuart~J Russell, and Thomas~L Griffiths.
\newblock Cognitive model priors for predicting human decisions.
\newblock In \emph{International conference on machine learning}, pages 5133--5141. PMLR, 2019.

\bibitem[Brophy(1984)]{brophy1984teacher}
Jere~E Brophy.
\newblock \emph{Teacher behavior and student achievement}.
\newblock Number~73. Institute for Research on Teaching, Michigan State University, 1984.

\bibitem[Bui et~al.(2022)Bui, Nhan, Dang, and Phung]{bui2022online}
Dien~Thi Bui, Thuy~Thi Nhan, Hue Thi~Thu Dang, and Trang Thi~Thu Phung.
\newblock Online learning experiences of secondary school students during covid-19--dataset from vietnam.
\newblock \emph{Data in Brief}, 45:\penalty0 108662, 2022.

\bibitem[Chamorro-Premuzic and Furnham(2003)]{chamorro2003personality}
Tomas Chamorro-Premuzic and Adrian Furnham.
\newblock Personality predicts academic performance: Evidence from two longitudinal university samples.
\newblock \emph{Journal of research in personality}, 37\penalty0 (4):\penalty0 319--338, 2003.

\bibitem[Chen et~al.(2018)Chen, Lu, Zheng, and Pian]{chen2018prerequisite}
Penghe Chen, Yu~Lu, Vincent~W Zheng, and Yang Pian.
\newblock Prerequisite-driven deep knowledge tracing.
\newblock In \emph{2018 IEEE International Conference on Data Mining (ICDM)}, pages 39--48. IEEE, 2018.

\bibitem[Chen et~al.(2023)Chen, Su, Zuo, Yang, Yuan, Qian, Chan, Qin, Lu, Xie, et~al.]{chen2023agentverse}
Weize Chen, Yusheng Su, Jingwei Zuo, Cheng Yang, Chenfei Yuan, Chen Qian, Chi-Min Chan, Yujia Qin, Yaxi Lu, Ruobing Xie, et~al.
\newblock Agentverse: Facilitating multi-agent collaboration and exploring emergent behaviors in agents.
\newblock \emph{arXiv preprint arXiv:2308.10848}, 2023.

\bibitem[Choi et~al.(2020)Choi, Lee, Shin, Cho, Park, Lee, Baek, Bae, Kim, and Heo]{choi2020ednet}
Youngduck Choi, Youngnam Lee, Dongmin Shin, Junghyun Cho, Seoyon Park, Seewoo Lee, Jineon Baek, Chan Bae, Byungsoo Kim, and Jaewe Heo.
\newblock Ednet: A large-scale hierarchical dataset in education.
\newblock In \emph{Artificial Intelligence in Education: 21st International Conference, AIED 2020, Ifrane, Morocco, July 6--10, 2020, Proceedings, Part II 21}, pages 69--73. Springer, 2020.

\bibitem[Cox(2023)]{cox2023use}
Samuel~Rhys Cox.
\newblock The use of multiple conversational agent interlocutors in learning.
\newblock \emph{arXiv preprint arXiv:2312.16534}, 2023.

\bibitem[Craig et~al.(2004)Craig, Graesser, Sullins, and Gholson]{AutoTutor}
Scotty~D. Craig, Arthur~C. Graesser, Jeremiah Sullins, and Barry Gholson.
\newblock Affect and learning: an exploratory look into the role of affect in learning with autotutor.
\newblock \emph{Journal of Educational Media}, 29, 2004.

\bibitem[Cziko(1989)]{cziko1989unpredictability}
Gary~A Cziko.
\newblock Unpredictability and indeterminism in human behavior: Arguments and implications for educational research.
\newblock \emph{Educational researcher}, 18\penalty0 (3):\penalty0 17--25, 1989.

\bibitem[Delgado et~al.(2021)Delgado, Origgi, Hasanpoor, Yu, Allessio, Arroyo, Lee, Betke, Woolf, and Bargal]{delgado2021student}
Kevin Delgado, Juan~Manuel Origgi, Tania Hasanpoor, Hao Yu, Danielle Allessio, Ivon Arroyo, William Lee, Margrit Betke, Beverly Woolf, and Sarah~Adel Bargal.
\newblock Student engagement dataset.
\newblock In \emph{Proceedings of the IEEE/CVF International Conference on Computer Vision}, pages 3628--3636, 2021.

\bibitem[Deng et~al.(2023)Deng, Gu, Zheng, Chen, Stevens, Wang, Sun, and Su]{deng2023mind2web}
Xiang Deng, Yu~Gu, Boyuan Zheng, Shijie Chen, Samuel Stevens, Boshi Wang, Huan Sun, and Yu~Su.
\newblock Mind2web: Towards a generalist agent for the web.
\newblock \emph{arXiv preprint arXiv:2306.06070}, 2023.

\bibitem[Deslauriers et~al.(2011)Deslauriers, Schelew, and Wieman]{impLearn}
Louis Deslauriers, Ellen Schelew, and Carl Wieman.
\newblock Improved learning in a large-enrollment physics class.
\newblock \emph{Science}, 332\penalty0 (6031):\penalty0 862--864, 2011.

\bibitem[Diaz-Piedra et~al.(2019)Diaz-Piedra, Rieiro, Cherino, Fuentes, Catena, and Di~Stasi]{diaz2019effects}
Carolina Diaz-Piedra, Hector Rieiro, Alberto Cherino, Luis~J Fuentes, Andres Catena, and Leandro~L Di~Stasi.
\newblock The effects of flight complexity on gaze entropy: An experimental study with fighter pilots.
\newblock \emph{Applied ergonomics}, 77:\penalty0 92--99, 2019.

\bibitem[D'Mello et~al.(2012)D'Mello, Olney, Williams, and Hays]{d2012gaze}
Sidney D'Mello, Andrew Olney, Claire Williams, and Patrick Hays.
\newblock Gaze tutor: A gaze-reactive intelligent tutoring system.
\newblock \emph{International Journal of human-computer studies}, 70\penalty0 (5):\penalty0 377--398, 2012.

\bibitem[Fan(2023)]{fan2023scb}
Yang Fan.
\newblock Scb-dataset: A dataset for detecting student classroom behavior.
\newblock \emph{arXiv preprint arXiv:2304.02488}, 2023.

\bibitem[Feng et~al.(2009)Feng, Heffernan, and Koedinger]{feng2009addressing}
Mingyu Feng, Neil Heffernan, and Kenneth Koedinger.
\newblock Addressing the assessment challenge with an online system that tutors as it assesses.
\newblock \emph{User modeling and user-adapted interaction}, 19:\penalty0 243--266, 2009.

\bibitem[Furnham et~al.(2003)Furnham, Chamorro-Premuzic, and McDougall]{furnham2003personality}
Adrian Furnham, Tomas Chamorro-Premuzic, and Fiona McDougall.
\newblock Personality, cognitive ability, and beliefs about intelligence as predictors of academic performance.
\newblock \emph{Learning and individual Differences}, 14\penalty0 (1):\penalty0 47--64, 2003.

\bibitem[Gao et~al.(2023)Gao, Lan, Lu, Mao, Piao, Wang, Jin, and Li]{gao2023s3}
Chen Gao, Xiaochong Lan, Zhihong Lu, Jinzhu Mao, Jinghua Piao, Huandong Wang, Depeng Jin, and Yong Li.
\newblock S3: Social-network simulation system with large language model-empowered agents, 2023.

\bibitem[Gong et~al.(2023)Gong, Huang, Ma, Vo, Durante, Noda, Zheng, Zhu, Terzopoulos, Fei-Fei, et~al.]{gong2023mindagent}
Ran Gong, Qiuyuan Huang, Xiaojian Ma, Hoi Vo, Zane Durante, Yusuke Noda, Zilong Zheng, Song-Chun Zhu, Demetri Terzopoulos, Li~Fei-Fei, et~al.
\newblock Mindagent: Emergent gaming interaction.
\newblock \emph{arXiv preprint arXiv:2309.09971}, 2023.

\bibitem[Gottlieb et~al.(2013)Gottlieb, Oudeyer, Lopes, and Baranes]{gottlieb2013information}
Jacqueline Gottlieb, Pierre-Yves Oudeyer, Manuel Lopes, and Adrien Baranes.
\newblock Information-seeking, curiosity, and attention: computational and neural mechanisms.
\newblock \emph{Trends in cognitive sciences}, 17\penalty0 (11):\penalty0 585--593, 2013.

\bibitem[Guo and Agichtein(2010)]{guo2010towards}
Qi~Guo and Eugene Agichtein.
\newblock Towards predicting web searcher gaze position from mouse movements.
\newblock In \emph{CHI'10 extended abstracts on human factors in computing systems}, pages 3601--3606. 2010.

\bibitem[Harandi(2015)]{harandi2015effects}
Safiyeh~Rajaee Harandi.
\newblock Effects of e-learning on students’ motivation.
\newblock \emph{Procedia-Social and Behavioral Sciences}, 181:\penalty0 423--430, 2015.

\bibitem[Hasan et~al.(2021)Hasan, Palaniappan, Mahmood, Abbas, and Sarker]{hasan2021dataset}
Raza Hasan, Sellappan Palaniappan, Salman Mahmood, Ali Abbas, and Kamal~Uddin Sarker.
\newblock Dataset of students’ performance using student information system, moodle and the mobile application “edify”.
\newblock \emph{Data}, 6\penalty0 (11):\penalty0 110, 2021.

\bibitem[Hussain et~al.(2019)Hussain, Zhu, Zhang, Abidi, and Ali]{hussain2019using}
Mushtaq Hussain, Wenhao Zhu, Wu~Zhang, Syed Muhammad~Raza Abidi, and Sadaqat Ali.
\newblock Using machine learning to predict student difficulties from learning session data.
\newblock \emph{Artificial Intelligence Review}, 52:\penalty0 381--407, 2019.

\bibitem[Jeon and Lee(2023)]{jeon2023large}
Jaeho Jeon and Seongyong Lee.
\newblock Large language models in education: A focus on the complementary relationship between human teachers and chatgpt.
\newblock \emph{Education and Information Technologies}, pages 1--20, 2023.

\bibitem[Jin et~al.(2023)Jin, Chen, Ye, Yang, Feng, Zhang, Yu, and Wang]{jin2023lending}
Jiarui Jin, Xianyu Chen, Fanghua Ye, Mengyue Yang, Yue Feng, Weinan Zhang, Yong Yu, and Jun Wang.
\newblock Lending interaction wings to recommender systems with conversational agents.
\newblock \emph{arXiv preprint arXiv:2310.04230}, 2023.

\bibitem[Jinxin et~al.(2023)Jinxin, Jiabao, Yilei, Xingjiao, Jiawen, and Liang]{jinxin2023cgmi}
Shi Jinxin, Zhao Jiabao, Wang Yilei, Wu~Xingjiao, Li~Jiawen, and He~Liang.
\newblock Cgmi: Configurable general multi-agent interaction framework.
\newblock \emph{arXiv preprint arXiv:2308.12503}, 2023.

\bibitem[Karemera et~al.(2003)Karemera, Reuben, and Sillah]{karemera2003effects}
David Karemera, Lucy~J Reuben, and Marion~R Sillah.
\newblock The effects of academic environment and background characteristics on student satisfaction and performance: The case of south carolina state university's school of business.
\newblock \emph{College Student Journal}, 37\penalty0 (2):\penalty0 298--309, 2003.

\bibitem[Kaur et~al.(2018)Kaur, Mustafa, Mehta, and Dhall]{kaur2018prediction}
Amanjot Kaur, Aamir Mustafa, Love Mehta, and Abhinav Dhall.
\newblock Prediction and localization of student engagement in the wild.
\newblock In \emph{2018 Digital Image Computing: Techniques and Applications (DICTA)}, pages 1--8. IEEE, 2018.

\bibitem[Kosel et~al.(2021)Kosel, Holzberger, and Seidel]{kosel2021identifying}
Christian Kosel, Doris Holzberger, and Tina Seidel.
\newblock Identifying expert and novice visual scanpath patterns and their relationship to assessing learning-relevant student characteristics.
\newblock In \emph{Frontiers in Education}, volume~5, page 612175. Frontiers Media SA, 2021.

\bibitem[Kung et~al.(2023)Kung, Cheatham, Medenilla, Sillos, De~Leon, Elepa{\~n}o, Madriaga, Aggabao, Diaz-Candido, Maningo, et~al.]{kung2023performance}
Tiffany~H Kung, Morgan Cheatham, Arielle Medenilla, Czarina Sillos, Lorie De~Leon, Camille Elepa{\~n}o, Maria Madriaga, Rimel Aggabao, Giezel Diaz-Candido, James Maningo, et~al.
\newblock Performance of chatgpt on usmle: Potential for ai-assisted medical education using large language models.
\newblock \emph{PLoS digital health}, 2\penalty0 (2):\penalty0 e0000198, 2023.

\bibitem[Kuzilek et~al.(2017)Kuzilek, Hlosta, and Zdrahal]{kuzilek2017open}
Jakub Kuzilek, Martin Hlosta, and Zdenek Zdrahal.
\newblock Open university learning analytics dataset.
\newblock \emph{Scientific data}, 4\penalty0 (1):\penalty0 1--8, 2017.

\bibitem[Kwok et~al.(2018)]{kwok2018understanding}
Cho~Ki Kwok et~al.
\newblock Understanding user engagement level during tasks via facial responses, eye gaze and mouse movements.
\newblock 2018.

\bibitem[Lau and Roeser(2002)]{lau2002cognitive}
Shun Lau and Robert~W Roeser.
\newblock Cognitive abilities and motivational processes in high school students' situational engagement and achievement in science.
\newblock \emph{Educational Assessment}, 8\penalty0 (2):\penalty0 139--162, 2002.

\bibitem[Lee et~al.(2021)Lee, Tzeng, Huang, and Su]{lee2021prediction}
Chia-An Lee, Jian-Wei Tzeng, Nen-Fu Huang, and Yu-Sheng Su.
\newblock Prediction of student performance in massive open online courses using deep learning system based on learning behaviors.
\newblock \emph{Educational Technology \& Society}, 24\penalty0 (3):\penalty0 130--146, 2021.

\bibitem[Lei et~al.(2018)Lei, Cui, and Zhou]{lei2018relationships}
Hao Lei, Yunhuo Cui, and Wenye Zhou.
\newblock Relationships between student engagement and academic achievement: A meta-analysis.
\newblock \emph{Social Behavior and Personality: an international journal}, 46\penalty0 (3):\penalty0 517--528, 2018.

\bibitem[Li et~al.(2023)Li, Hammoud, Itani, Khizbullin, and Ghanem]{li2023camel}
Guohao Li, Hasan Abed Al~Kader Hammoud, Hani Itani, Dmitrii Khizbullin, and Bernard Ghanem.
\newblock Camel: Communicative agents for" mind" exploration of large language model society.
\newblock In \emph{Thirty-seventh Conference on Neural Information Processing Systems}, 2023.

\bibitem[Lin et~al.(2023)Lin, Fu, Yang, Brahman, Huang, Bhagavatula, Ammanabrolu, Choi, and Ren]{lin2023swiftsage}
Bill~Yuchen Lin, Yicheng Fu, Karina Yang, Faeze Brahman, Shiyu Huang, Chandra Bhagavatula, Prithviraj Ammanabrolu, Yejin Choi, and Xiang Ren.
\newblock Swiftsage: A generative agent with fast and slow thinking for complex interactive tasks.
\newblock \emph{arXiv preprint arXiv:2305.17390}, 2023.

\bibitem[Liu et~al.(2017)Liu, McKelroy, Corliss, and Carrigan]{liu2017investigating}
Min Liu, Emily McKelroy, Stephanie~B Corliss, and Jamison Carrigan.
\newblock Investigating the effect of an adaptive learning intervention on students’ learning.
\newblock \emph{Educational technology research and development}, 65:\penalty0 1605--1625, 2017.

\bibitem[Liu et~al.(2019)Liu, Huang, Yin, Chen, Xiong, Su, and Hu]{liu2019ekt}
Qi~Liu, Zhenya Huang, Yu~Yin, Enhong Chen, Hui Xiong, Yu~Su, and Guoping Hu.
\newblock Ekt: Exercise-aware knowledge tracing for student performance prediction.
\newblock \emph{IEEE Transactions on Knowledge and Data Engineering}, 33\penalty0 (1):\penalty0 100--115, 2019.

\bibitem[Liu et~al.(2023)Liu, Liu, Guo, Chen, Huang, Zhao, Tang, Luo, and Weng]{liu2023xes3g5m}
Zitao Liu, Qiongqiong Liu, Teng Guo, Jiahao Chen, Shuyan Huang, Xiangyu Zhao, Jiliang Tang, Weiqi Luo, and Jian Weng.
\newblock Xes3g5m: A knowledge tracing benchmark dataset with auxiliary information.
\newblock In \emph{Thirty-seventh Conference on Neural Information Processing Systems Datasets and Benchmarks Track}, 2023.

\bibitem[Mai et~al.(2022)Mai, Bezbradica, and Crane]{mai2022learning}
Tai~Tan Mai, Marija Bezbradica, and Martin Crane.
\newblock Learning behaviours data in programming education: Community analysis and outcome prediction with cleaned data.
\newblock \emph{Future Generation Computer Systems}, 127:\penalty0 42--55, 2022.

\bibitem[Markel et~al.(2023)Markel, Opferman, Landay, and Piech]{GPTeach}
Julia~M Markel, Steven~G Opferman, James~A Landay, and Chris Piech.
\newblock {GPTeach: Interactive TA Training with GPT Based Students}.
\newblock 2023.

\bibitem[Mart{\'\i}n et~al.(2015)Mart{\'\i}n, G{\'e}rtrudix, Urquiza-Fuentes, and Haya]{martin2015student}
Estefan{\'\i}a Mart{\'\i}n, Manuel G{\'e}rtrudix, Jaime Urquiza-Fuentes, and Pablo~A Haya.
\newblock Student activity and profile datasets from an online video-based collaborative learning experience.
\newblock \emph{British Journal of Educational Technology}, 46\penalty0 (5):\penalty0 993--998, 2015.

\bibitem[Massaro et~al.(2012)Massaro, Savazzi, Di~Dio, Freedberg, Gallese, Gilli, and Marchetti]{massaro2012art}
Davide Massaro, Federica Savazzi, Cinzia Di~Dio, David Freedberg, Vittorio Gallese, Gabriella Gilli, and Antonella Marchetti.
\newblock When art moves the eyes: a behavioral and eye-tracking study.
\newblock \emph{PloS one}, 7\penalty0 (5):\penalty0 e37285, 2012.

\bibitem[Matelsky et~al.(2023)Matelsky, Parodi, Liu, Lange, and Kording]{matelsky2023large}
Jordan~K Matelsky, Felipe Parodi, Tony Liu, Richard~D Lange, and Konrad~P Kording.
\newblock A large language model-assisted education tool to provide feedback on open-ended responses.
\newblock \emph{arXiv preprint arXiv:2308.02439}, 2023.

\bibitem[Mayer(2009)]{MayerBook}
Richard~E. Mayer.
\newblock \emph{Multimedia Learning}.
\newblock Cambridge University Press, 2nd edition, 2009.
\newblock ISBN 0521514126.

\bibitem[McLoughlin(2001)]{mcloughlin2001inclusivity}
Catherine McLoughlin.
\newblock Inclusivity and alignment: Principles of pedagogy, task and assessment design for effective cross-cultural online learning.
\newblock \emph{Distance Education}, 22\penalty0 (1):\penalty0 7--29, 2001.

\bibitem[Minn et~al.(2018)Minn, Yu, Desmarais, Zhu, and Vie]{minn2018deep}
Sein Minn, Yi~Yu, Michel~C Desmarais, Feida Zhu, and Jill-Jenn Vie.
\newblock Deep knowledge tracing and dynamic student classification for knowledge tracing.
\newblock In \emph{2018 IEEE International conference on data mining (ICDM)}, pages 1182--1187. IEEE, 2018.

\bibitem[Nakayama et~al.(2021)Nakayama, Mutsuura, and Yamamoto]{nakayama2021impact}
Minoru Nakayama, Kouichi Mutsuura, and Hiroh Yamamoto.
\newblock Impact of learner’s characteristics and learning behaviour on learning performance during a fully online course.
\newblock \emph{Note taking activities in e-learning environments}, pages 15--36, 2021.

\bibitem[Pardos et~al.(2013)Pardos, Baker, San~Pedro, Gowda, and Gowda]{pardos2013affective}
Zachary~A Pardos, Ryan~SJD Baker, Maria~OCZ San~Pedro, Sujith~M Gowda, and Supreeth~M Gowda.
\newblock Affective states and state tests: Investigating how affect throughout the school year predicts end of year learning outcomes.
\newblock In \emph{Proceedings of the third international conference on learning analytics and knowledge}, pages 117--124, 2013.

\bibitem[Park et~al.(2023)Park, O'Brien, Cai, Morris, Liang, and Bernstein]{park2023generative}
Joon~Sung Park, Joseph O'Brien, Carrie~Jun Cai, Meredith~Ringel Morris, Percy Liang, and Michael~S Bernstein.
\newblock Generative agents: Interactive simulacra of human behavior.
\newblock In \emph{Proceedings of the 36th Annual ACM Symposium on User Interface Software and Technology}, pages 1--22, 2023.

\bibitem[Piech et~al.(2015)Piech, Bassen, Huang, Ganguli, Sahami, Guibas, and Sohl-Dickstein]{piech2015deep}
Chris Piech, Jonathan Bassen, Jonathan Huang, Surya Ganguli, Mehran Sahami, Leonidas~J Guibas, and Jascha Sohl-Dickstein.
\newblock Deep knowledge tracing.
\newblock \emph{Advances in neural information processing systems}, 28, 2015.

\bibitem[Pojen et~al.(2020)Pojen, Mingen, and Tzuyang]{JunyiOnlineLearningDataset}
Chen Pojen, Hsieh Mingen, and Tsai Tzuyang.
\newblock Junyi academy online learning activity dataset: A large-scale public online learning activity dataset from elementary to senior high school students.
\newblock \emph{Dataset available from https://www.kaggle.com/junyiacademy/learning-activity-public-dataset-by-junyi-academy}, 2020.

\bibitem[Resnick(2017)]{resnick2017toward}
Lauren~B Resnick.
\newblock Toward a cognitive theory of instruction.
\newblock In \emph{Learning and motivation in the classroom}, pages 5--38. Routledge, 2017.

\bibitem[Ruiz et~al.(2022)Ruiz, Yu, Allessio, Jalal, Joshi, Murray, Magee, Delgado, Ablavsky, Sclaroff, et~al.]{ruiz2022atl}
Nataniel Ruiz, Hao Yu, Danielle~A Allessio, Mona Jalal, Ajjen Joshi, Tom Murray, John~J Magee, Kevin~Manuel Delgado, Vitaly Ablavsky, Stan Sclaroff, et~al.
\newblock Atl-bp: a student engagement dataset and model for affect transfer learning for behavior prediction.
\newblock \emph{IEEE Transactions on Biometrics, Behavior, and Identity Science}, 2022.

\bibitem[Seidel et~al.(2021)Seidel, Schnitzler, Kosel, St{\"u}rmer, and Holzberger]{seidel2021student}
Tina Seidel, Katharina Schnitzler, Christian Kosel, Kathleen St{\"u}rmer, and Doris Holzberger.
\newblock Student characteristics in the eyes of teachers: Differences between novice and expert teachers in judgment accuracy, observed behavioral cues, and gaze.
\newblock \emph{Educational Psychology Review}, 33:\penalty0 69--89, 2021.

\bibitem[Stamper and Pardos(2016)]{stamper20162010}
John Stamper and Zachary~A Pardos.
\newblock The 2010 kdd cup competition dataset: Engaging the machine learning community in predictive learning analytics.
\newblock \emph{Journal of Learning Analytics}, 3\penalty0 (2):\penalty0 312--316, 2016.

\bibitem[Sun et~al.(2021)Sun, Wu, Zhao, He, Yu, Yan, and Luo]{sun2021student}
Bo~Sun, Yong Wu, Kaijie Zhao, Jun He, Lejun Yu, Huanqing Yan, and Ao~Luo.
\newblock Student class behavior dataset: a video dataset for recognizing, detecting, and captioning students’ behaviors in classroom scenes.
\newblock \emph{Neural Computing and Applications}, 33:\penalty0 8335--8354, 2021.

\bibitem[Syed et~al.(2020)Syed, Collins-Thompson, Bennett, Teng, Williams, Tay, and Iqbal]{Qgen}
Rohail Syed, Kevyn Collins-Thompson, Paul~N Bennett, Mengqiu Teng, Shane Williams, Dr~Wendy~W Tay, and Shamsi Iqbal.
\newblock Improving learning outcomes with gaze tracking and automatic question generation.
\newblock In \emph{Proceedings of The Web Conference 2020}, 2020.

\bibitem[Uscher-Pines et~al.(2018)Uscher-Pines, Schwartz, Ahmed, Zheteyeva, Meza, Baker, and Uzicanin]{uscher2018school}
Lori Uscher-Pines, Heather~L Schwartz, Faruque Ahmed, Yenlik Zheteyeva, Erika Meza, Garrett Baker, and Amra Uzicanin.
\newblock School practices to promote social distancing in k-12 schools: review of influenza pandemic policies and practices.
\newblock \emph{BMC public health}, 18\penalty0 (1):\penalty0 1--13, 2018.

\bibitem[Waheed et~al.(2020)Waheed, Hassan, Aljohani, Hardman, Alelyani, and Nawaz]{waheed2020predicting}
Hajra Waheed, Saeed-Ul Hassan, Naif~Radi Aljohani, Julie Hardman, Salem Alelyani, and Raheel Nawaz.
\newblock Predicting academic performance of students from vle big data using deep learning models.
\newblock \emph{Computers in Human behavior}, 104:\penalty0 106189, 2020.

\bibitem[Wang et~al.(2023)Wang, Zhang, Yang, Chen, Tang, Zhang, Chen, Lin, Song, Zhao, Xu, Dou, Wang, and Wen]{wang2023large}
Lei Wang, Jingsen Zhang, Hao Yang, Zhiyuan Chen, Jiakai Tang, Zeyu Zhang, Xu~Chen, Yankai Lin, Ruihua Song, Wayne~Xin Zhao, Jun Xu, Zhicheng Dou, Jun Wang, and Ji-Rong Wen.
\newblock When large language model based agent meets user behavior analysis: A novel user simulation paradigm, 2023.

\bibitem[Wang et~al.(2021)Wang, Lamb, Saveliev, Cameron, Zaykov, Hernandez-Lobato, Turner, Baraniuk, Barton, Jones, et~al.]{wang2021results}
Zichao Wang, Angus Lamb, Evgeny Saveliev, Pashmina Cameron, Jordan Zaykov, Jose~Miguel Hernandez-Lobato, Richard~E Turner, Richard~G Baraniuk, Craig Barton, Simon~Peyton Jones, et~al.
\newblock Results and insights from diagnostic questions: The neurips 2020 education challenge.
\newblock In \emph{NeurIPS 2020 Competition and Demonstration Track}, pages 191--205. PMLR, 2021.

\bibitem[Wei et~al.(2022)Wei, Wang, Schuurmans, Bosma, Xia, Chi, Le, Zhou, et~al.]{wei2022chain}
Jason Wei, Xuezhi Wang, Dale Schuurmans, Maarten Bosma, Fei Xia, Ed~Chi, Quoc~V Le, Denny Zhou, et~al.
\newblock Chain-of-thought prompting elicits reasoning in large language models.
\newblock \emph{Advances in Neural Information Processing Systems}, 35:\penalty0 24824--24837, 2022.

\bibitem[Xiong et~al.(2016)Xiong, Zhao, Van~Inwegen, and Beck]{xiong2016going}
Xiaolu Xiong, Siyuan Zhao, Eric~G Van~Inwegen, and Joseph~E Beck.
\newblock Going deeper with deep knowledge tracing.
\newblock \emph{International Educational Data Mining Society}, 2016.

\bibitem[Xu et~al.(2017)Xu, Moon, and Van Der~Schaar]{xu2017machine}
Jie Xu, Kyeong~Ho Moon, and Mihaela Van Der~Schaar.
\newblock A machine learning approach for tracking and predicting student performance in degree programs.
\newblock \emph{IEEE Journal of Selected Topics in Signal Processing}, 11\penalty0 (5):\penalty0 742--753, 2017.

\bibitem[Xu and Zhang(2023)]{xu2023leveraging}
Songlin Xu and Xinyu Zhang.
\newblock Leveraging generative artificial intelligence to simulate student learning behavior.
\newblock \emph{arXiv preprint arXiv:2310.19206}, 2023.

\bibitem[Xu et~al.(2023)Xu, Hu, Wang, and Zhang]{xu2023peer}
Songlin Xu, Dongyin Hu, Ru~Wang, and Xinyu Zhang.
\newblock Peer attention enhances student learning.
\newblock \emph{arXiv e-prints}, pages arXiv--2312, 2023.

\bibitem[Yao et~al.(2022)Yao, Chen, Yang, and Narasimhan]{yao2022webshop}
Shunyu Yao, Howard Chen, John Yang, and Karthik Narasimhan.
\newblock Webshop: Towards scalable real-world web interaction with grounded language agents.
\newblock \emph{Advances in Neural Information Processing Systems}, 35:\penalty0 20744--20757, 2022.

\bibitem[Zhang et~al.(2023)Zhang, Sheng, Chen, Li, Deng, Wang, and Chua]{zhang2023generative}
An~Zhang, Leheng Sheng, Yuxin Chen, Hao Li, Yang Deng, Xiang Wang, and Tat-Seng Chua.
\newblock On generative agents in recommendation.
\newblock \emph{arXiv preprint arXiv:2310.10108}, 2023.

\bibitem[Zhang et~al.(2020)Zhang, Li, Liu, Cao, and Liu]{zhang2020data}
Zhaoli Zhang, Zhenhua Li, Hai Liu, Taihe Cao, and Sannyuya Liu.
\newblock Data-driven online learning engagement detection via facial expression and mouse behavior recognition technology.
\newblock \emph{Journal of Educational Computing Research}, 58\penalty0 (1):\penalty0 63--86, 2020.

\bibitem[Zhao et~al.(2021)Zhao, Xu, and Thille]{zhao2021end}
Jinjin Zhao, Weijie Xu, and Candace Thille.
\newblock End-to-end question generation to assist formative assessment design for conceptual knowledge learning.
\newblock 2021.

\bibitem[Zhu et~al.(2023)Zhu, Shi, Song, and Cai]{zhu2023integrating}
Rongrong Zhu, Liang Shi, Yunpeng Song, and ZhongMin Cai.
\newblock Integrating gaze and mouse via joint cross-attention fusion net for students' activity recognition in e-learning.
\newblock \emph{Proceedings of the ACM on Interactive, Mobile, Wearable and Ubiquitous Technologies}, 7\penalty0 (3):\penalty0 1--35, 2023.

\end{thebibliography}
\bibliographystyle{plainnat}


\onecolumn
\section{Supplementary Material}


\subsection{\textbf{EduAgent310} dataset}


Here are the measurements we performed to get cognitive states.
\textbf{Workload} is represented by gaze stationary entropy in specific duration according to \cite{diaz2019effects}. \textbf{Curiosity} is represented by gaze transition entropy according to \cite{gottlieb2013information}. For each second, \textbf{valid focus} is denoted as 1 if students' gaze falls into any AOIs on slides. Otherwise, it is 0. \textbf{Course following} is denoted as 1 if gaze falling AOI is the same as the AOI that the teacher is just talking about (course pacing AOI). Otherwise, it is 0. \textbf{Engagement} is denoted as 0 if the student face is not detected by the web camera. Otherwise, it is 1. \textbf{Confusion} is denoted as 1 if students click the mouse to report their confusion. Otherwise, it is 0. All cognitive states are first calculated within each second and then get averaged during specific transcripts so all states are continuous values.

\begin{figure*}
\centering
\includegraphics[width=1\linewidth]{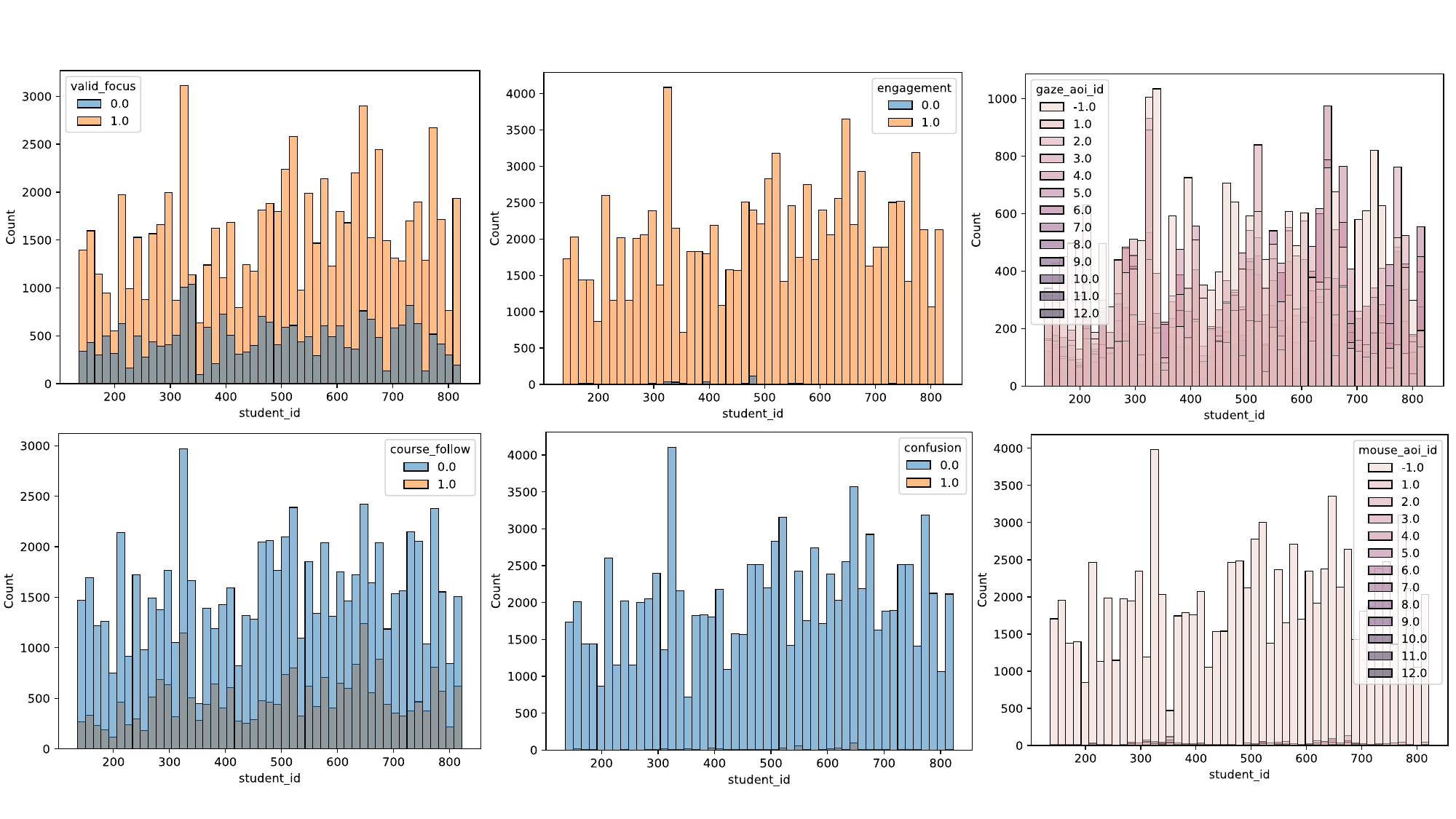}
\caption{Data distribution in \textbf{EduAgent310}.
}
\label{appendix fig: dataset 1 distribution 1}
\end{figure*}

\begin{figure*}
\centering
\includegraphics[width=1\linewidth]{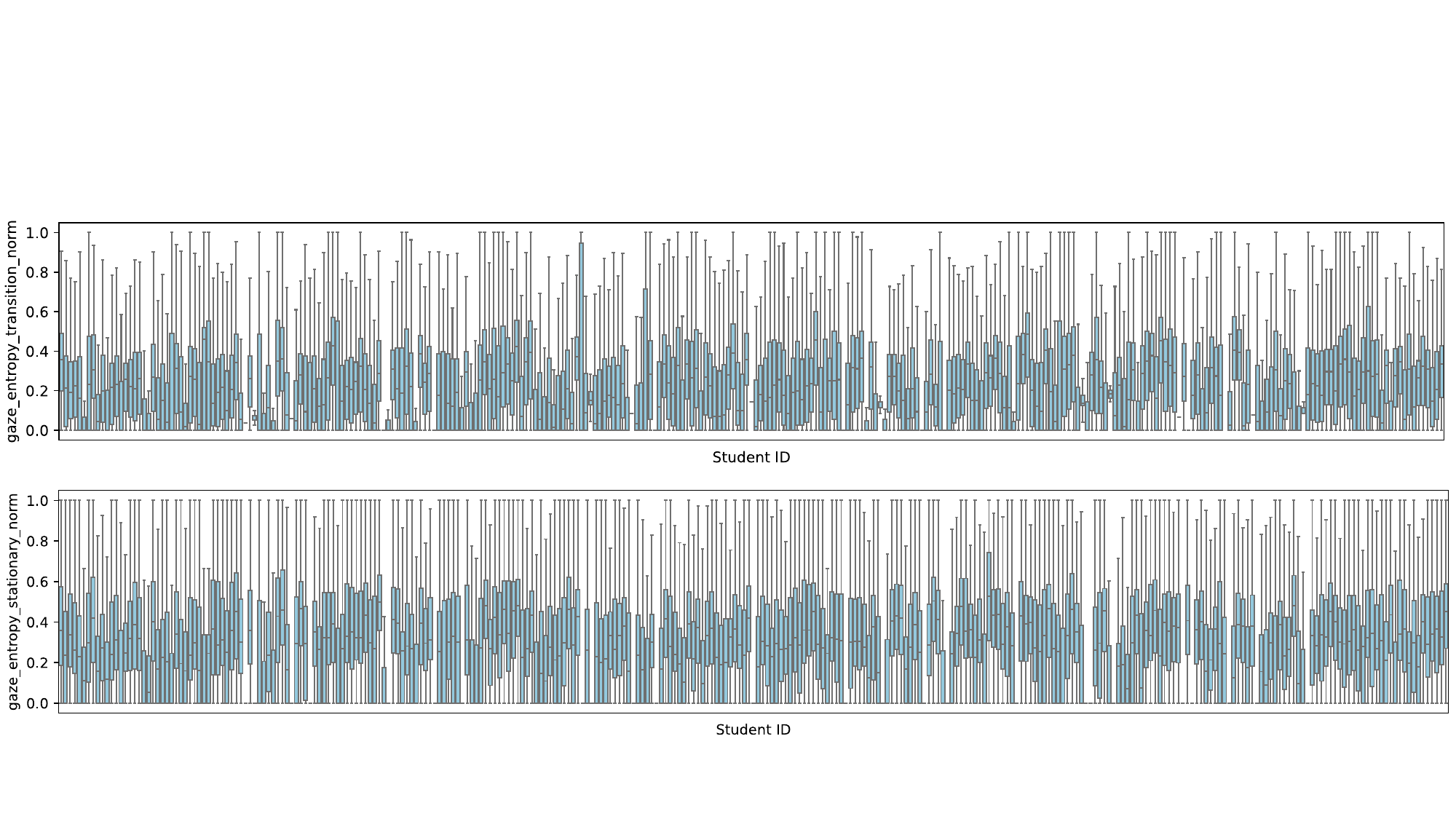}
\caption{Distribution of gaze stationary entropy (used to represent workload) and transition entropy (used to represent curiosity) in \textbf{EduAgent310} dataset.
}
\label{appendix fig: dataset 1 distribution 2}
\end{figure*}

Dataset distribution is depicted in Fig. \ref{appendix fig: dataset 1 distribution 1} and Fig. \ref{appendix fig: dataset 1 distribution 2}.

\subsection{\textbf{EduAgent705} dataset}

\begin{figure*}
\centering
\includegraphics[width=1\linewidth]{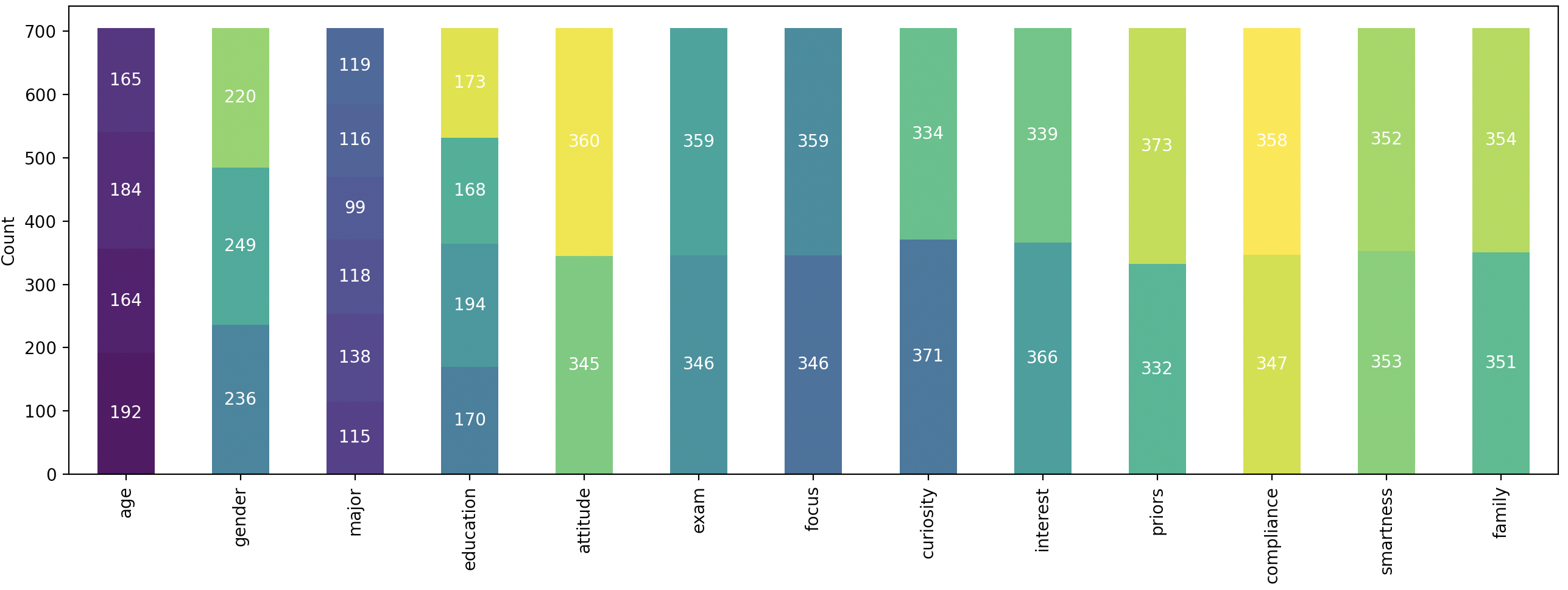}
\caption{Distribution of each kind of persona in \textbf{EduAgent705} dataset.
}
\label{appendix fig: dataset 2 distribution}
\end{figure*}

\begin{figure*}
\centering
\includegraphics[width=1\linewidth]{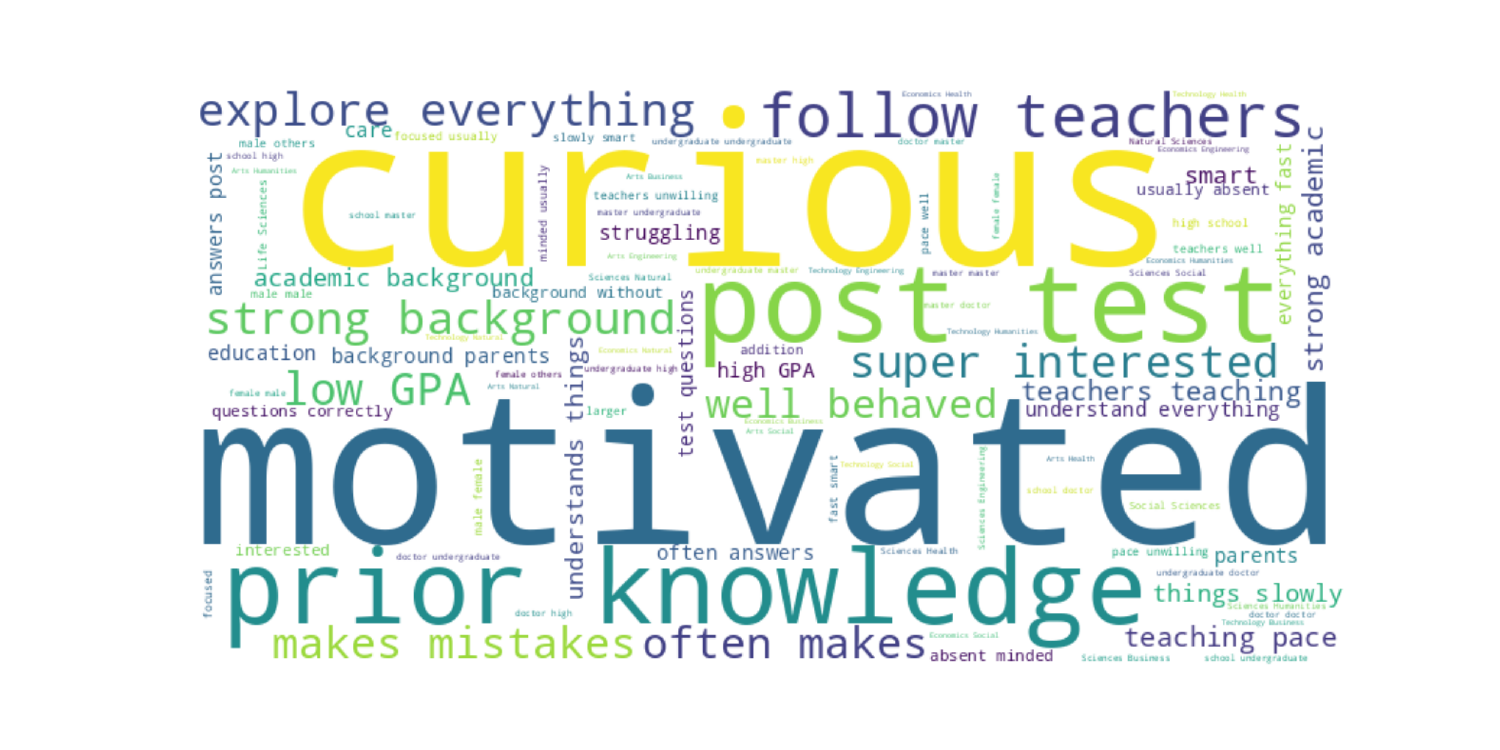}
\caption{Word cloud of personas of all agents in \textbf{EduAgent705} dataset.
}
\label{appendix fig: word cloud}
\end{figure*}

Dataset distribution is depicted in Fig. \ref{appendix fig: dataset 2 distribution} and details of personas are depicted in Table. \ref{appendix table:demo virtual student}. The word cloud figure that contains all personas in the dataset is depicted in Fig. \ref{appendix fig: word cloud}.

\subsection{Additional information of experiment 1}

Here we describe all metrics used in the first experiment in detail.

\textbf{Gaze/Motor}: As depicted in \ref{sec: framework}, the actions for gaze/motor are the simulated AOI ID on each slide. We compare the spatial \textbf{AOI distance} of the AOI center point location between agents and corresponding real students, serving as the metric. Note that all coordinates and AOI locations have been normalized into the range $[0,1]$ by adapting different students' screen size. The reason why we use AOI distance instead of 
AOI accuracy (whether agent AOI and real student AOI are exactly the same) is that: First, closeby AOIs are acceptable even if they are not the same considering the potential errors caused by gaze tracking techniques. Additionally, we do not use Top-N accuracy because different slides have different number of AOIs (usually ranging from 5 to 12), making it not a general comparison across slides. Finally, our ultimate goal is still to simulate the focused location on slides where students are watching or moving the mouse so distance (continuous value) is a better metric to measure location difference compared with accuracy (categorical value).

\textbf{Cognitive States}: We use Mean Absolute Error (MAE) between the simulated agents' cognitive states (normalized to 1) and the ground truth as metrics. 

\textbf{Question Answering}: We use \textit{choice similarity} and \textit{accuracy similarity} to quantify the answer choice difference and answer accuracy difference between agents and real students. Specifically, if a simulated agent and the real student make the same choice, then the choice similarity is 1 regardless of their choices are wrong or correct. Otherwise the choice similarity is 0. Whereas the accuracy similarity is 1 only when the agents' accuracy and real students' accuracy are the same by comparing with the correct question answers respectively. Otherwise the accuracy similarity is 0. Finally, we calculate the average results for both metrics.

\begin{figure*}
\centering
\includegraphics[width=1\linewidth]{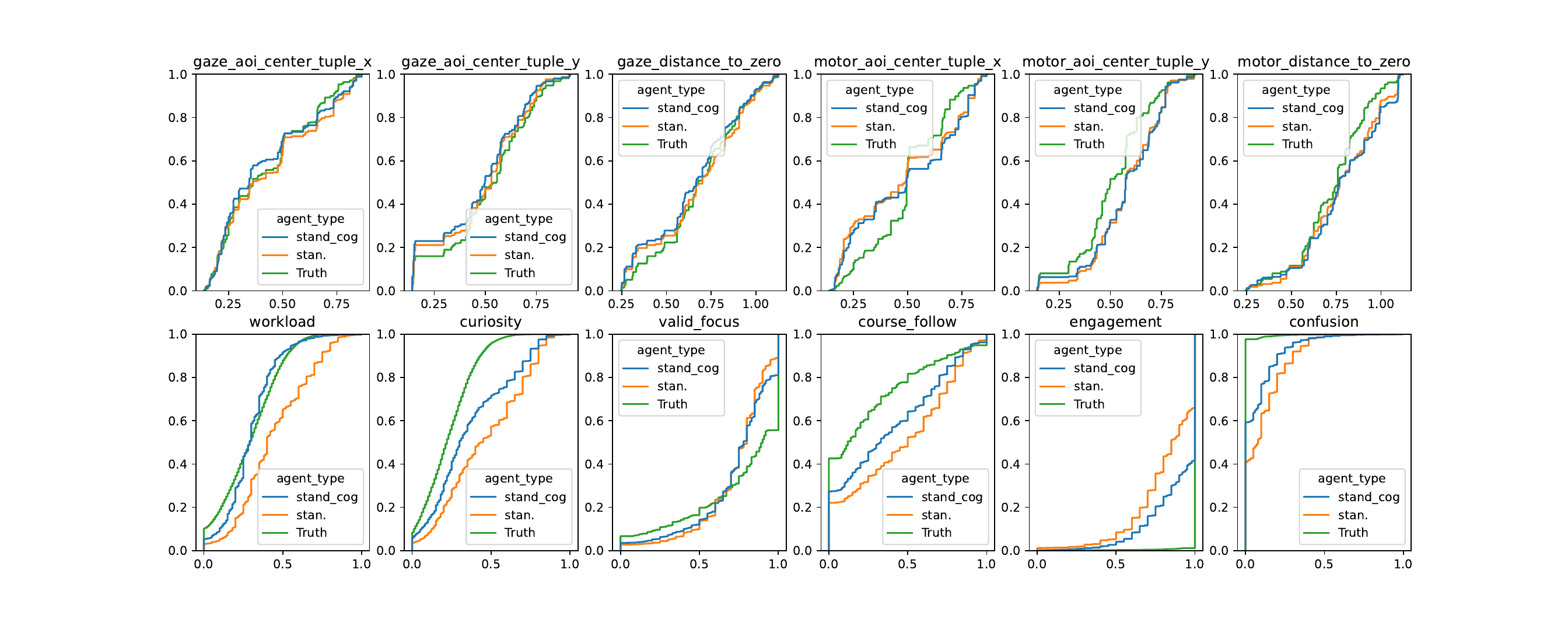}
\caption{CDF (Cumulative Distribution Function) plots among all metrics by comparing standard prompt (stan.) with the prompt integrating cognitive prior knowledge (standard cog) in the first experiment.
}
\label{appendix fig: cdf_strategy}
\end{figure*}

\begin{figure*}
\centering
\includegraphics[width=1\linewidth]{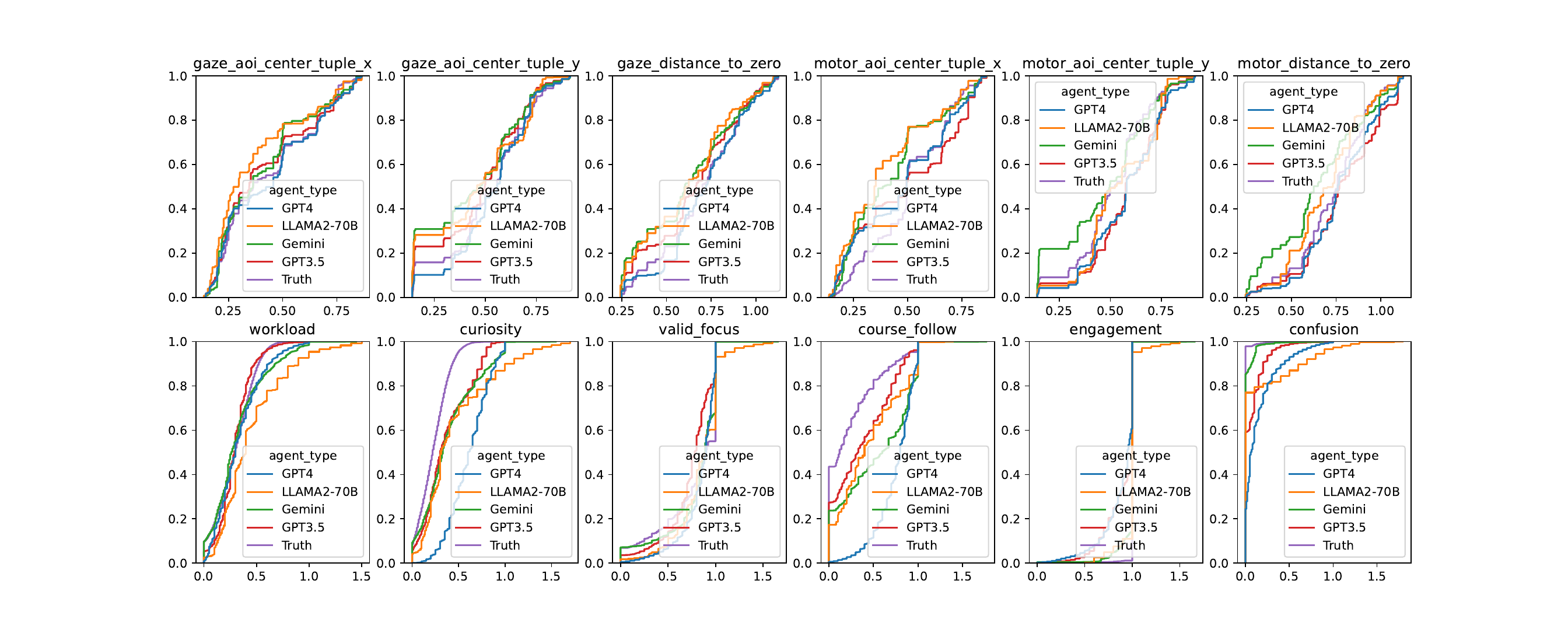}
\caption{CDF (Cumulative Distribution Function) plots among all metrics compared with different foundation models in the first experiment.
}
\label{appendix fig: cdf_gpt_type}
\end{figure*}

\begin{figure*}
\centering
\includegraphics[width=1\linewidth]{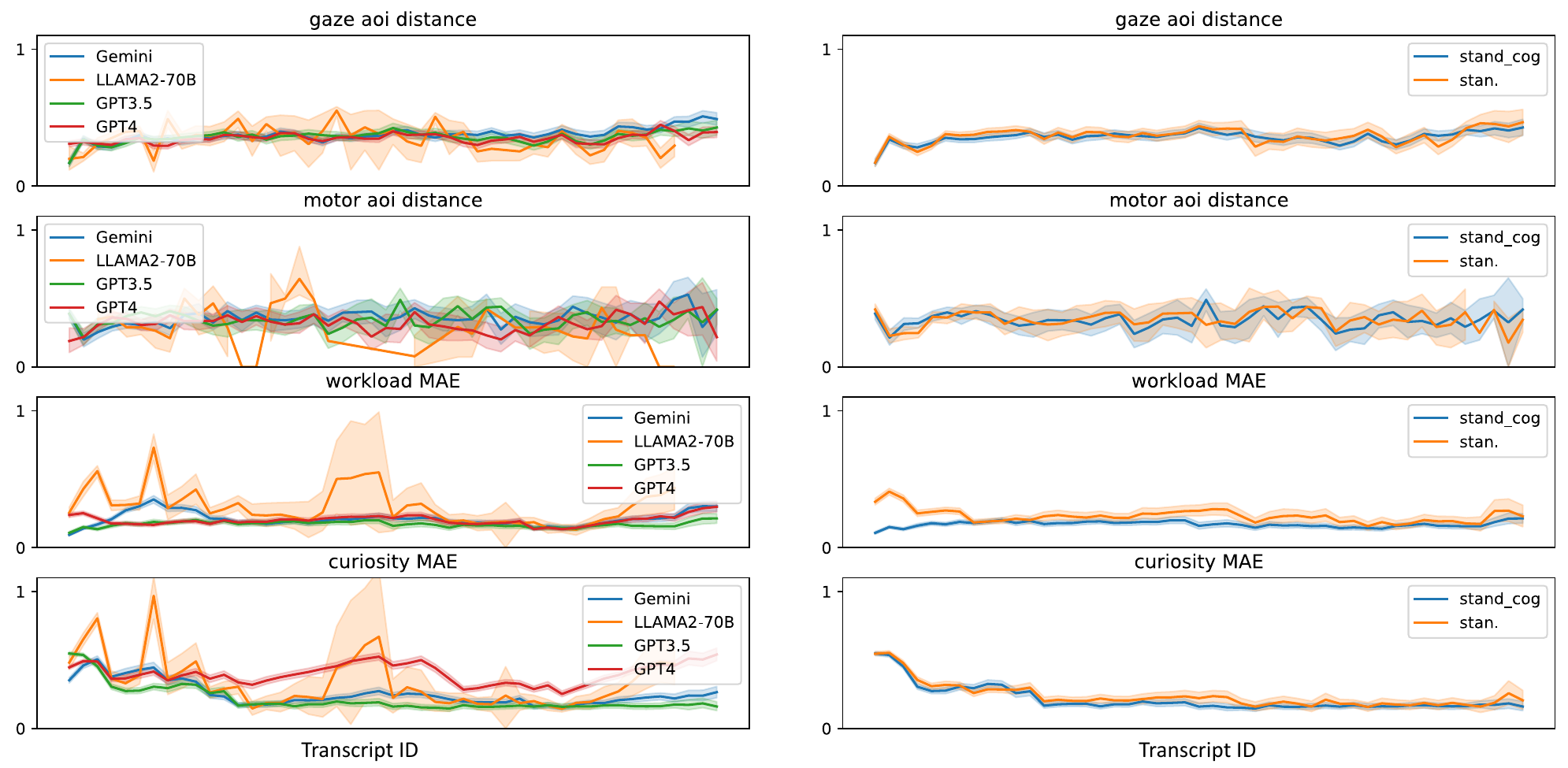}
\caption{Simulation performance that changes with transcript ID by comparing different foundation models and by comparing standard prompt (stan.) with the prompt integrating cognitive prior knowledge (standard cog) in the first experiment.
}
\label{appendix fig: transcript curve metrics 1}
\end{figure*}

\begin{figure*}
\centering
\includegraphics[width=1\linewidth]{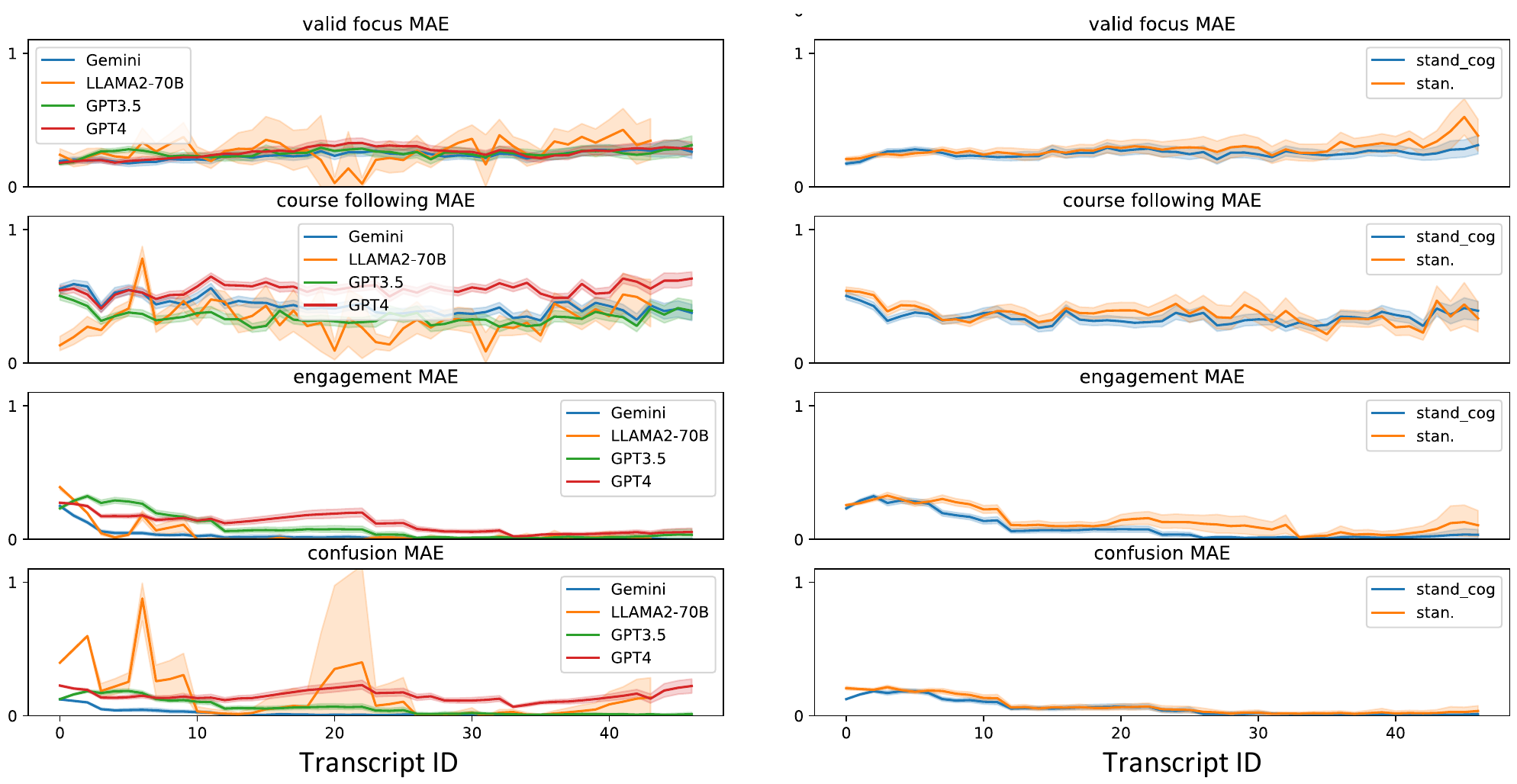}
\caption{Simulation performance that changes with transcript ID by comparing different foundation models and by comparing standard prompt (stan.) with the prompt integrating cognitive prior knowledge (standard cog) in the first experiment.
}
\label{appendix fig: transcript curve metrics 2}
\end{figure*}

\begin{figure*}
\centering
\includegraphics[width=1\linewidth]{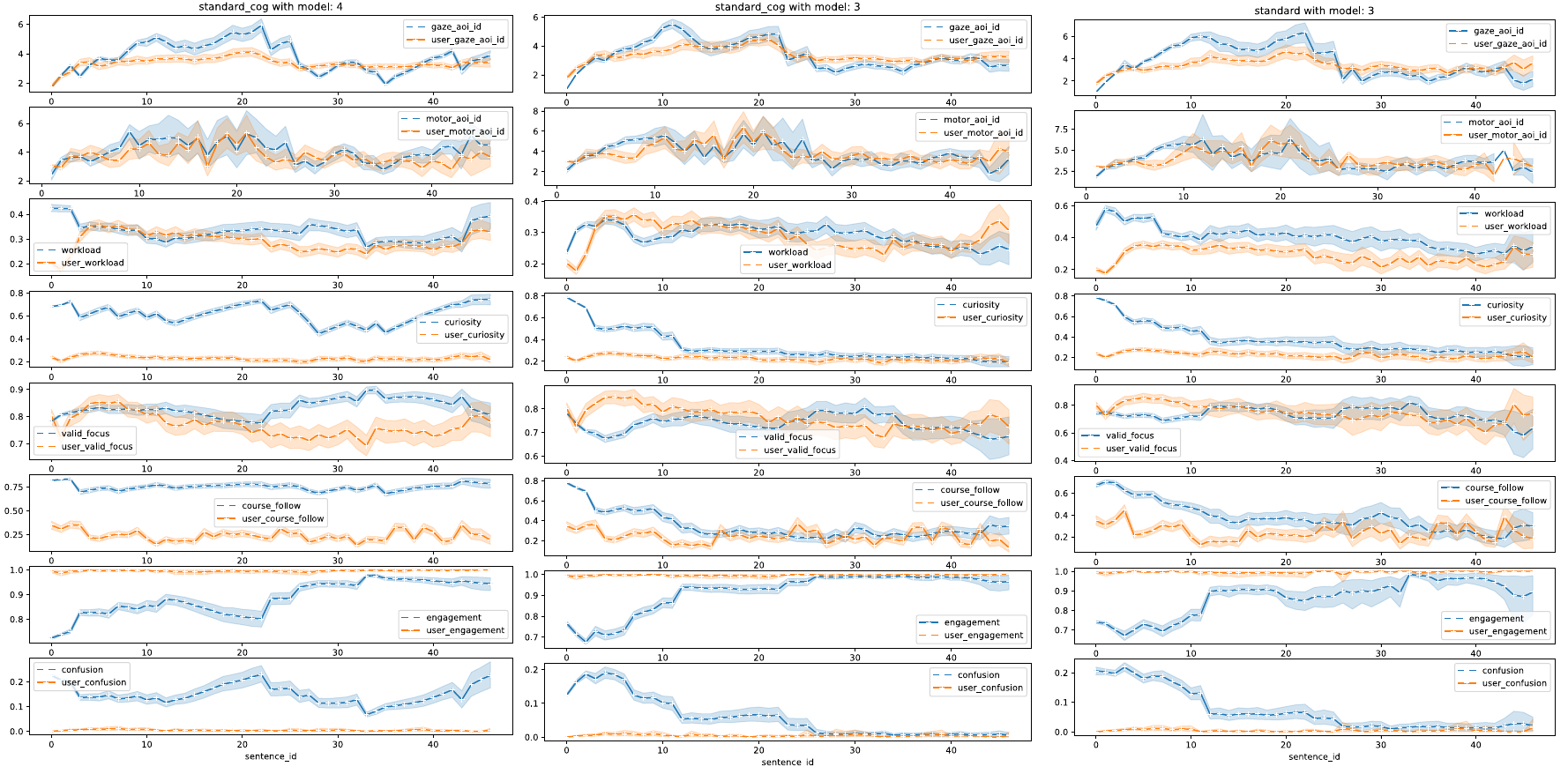}
\caption{Simulation performance that changes with transcript ID (sentence ID in the figure) by comparing different foundation models and different prompts in the first experiment. Model 3 refers to GPT-3.5 and model 4 refers to GPT-4. Standard cog uses our cognitive priors. Blue curves are agents simulation behaviors and orange curves are real students behaviors as ground truth.
}
\label{appendix fig: truth curve metrics}
\end{figure*}

\begin{figure*}
\centering
\includegraphics[width=1\linewidth]{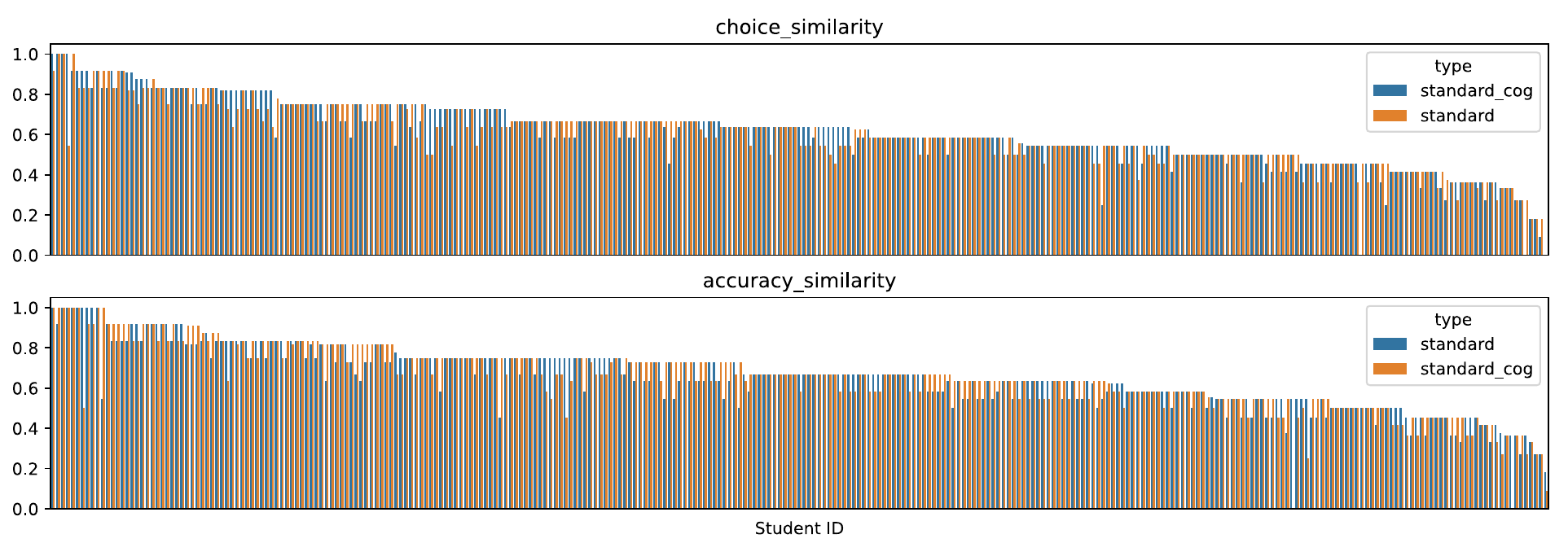}
\caption{Simulation performance in question answering for different students with different prompts in the first experiment. Standard cog uses our cognitive priors. 
}
\label{appendix fig: question ind compare}
\end{figure*}

\begin{figure*}
\centering
\includegraphics[width=1\linewidth]{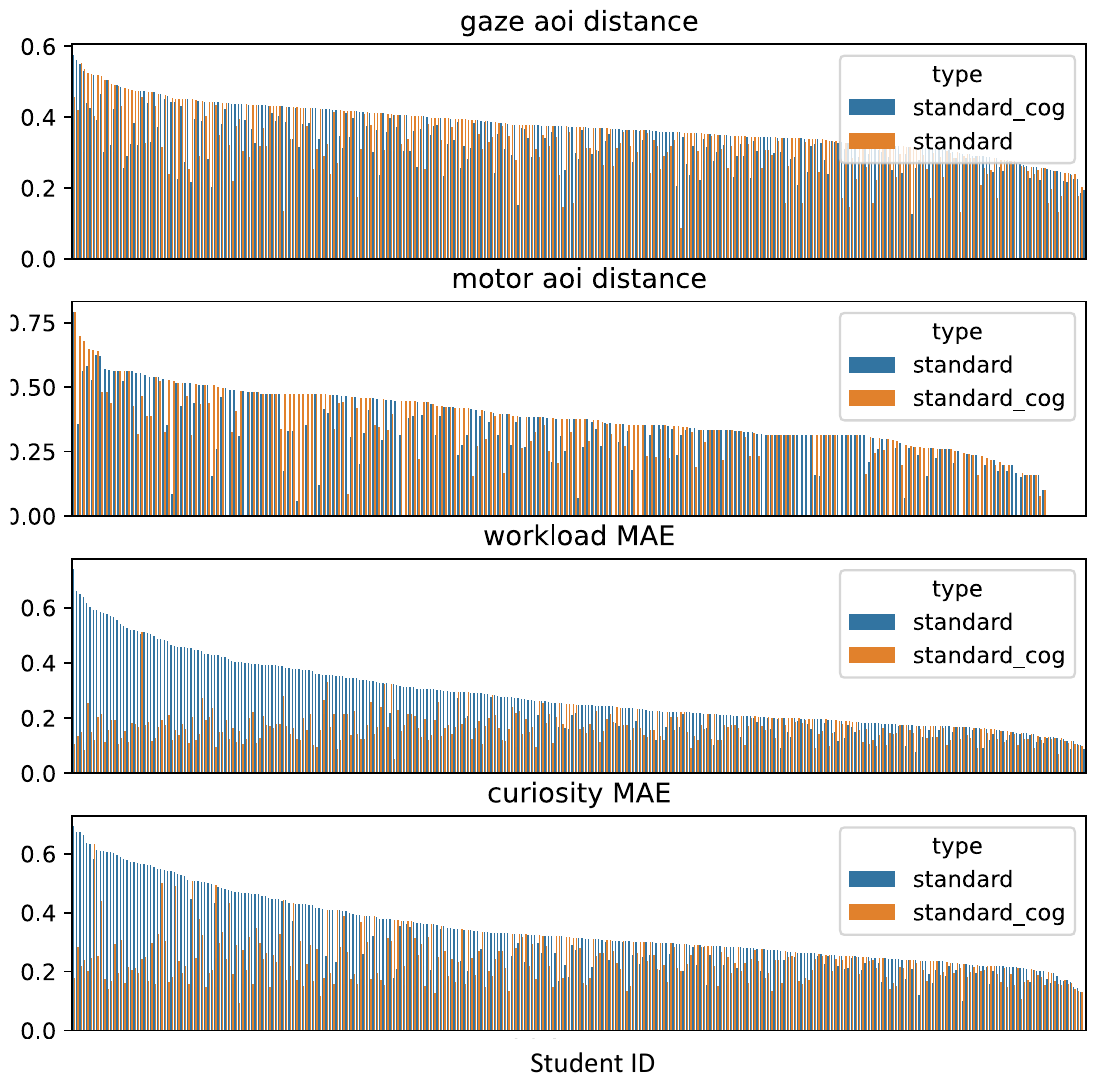}
\caption{Simulation performance for different students by comparing different prompts in the first experiment. Standard cog uses our cognitive priors. 
}
\label{appendix fig: dur ind compare 0}
\end{figure*}

\begin{figure*}
\centering
\includegraphics[width=1\linewidth]{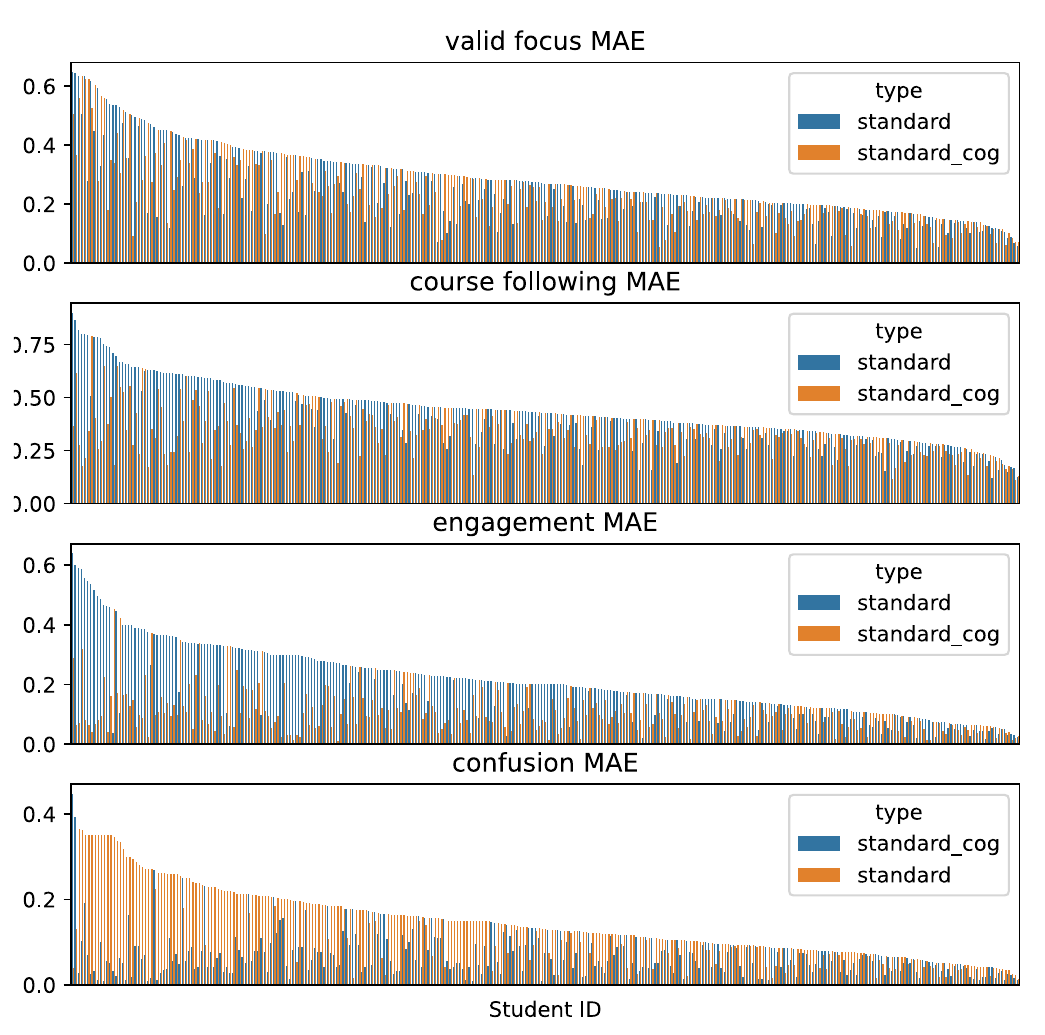}
\caption{Simulation performance for different students by comparing different prompts in the first experiment. Standard cog uses our cognitive priors. 
}
\label{appendix fig: dur ind compare 1}
\end{figure*}

\begin{figure*}
\centering
\includegraphics[width=1\linewidth]{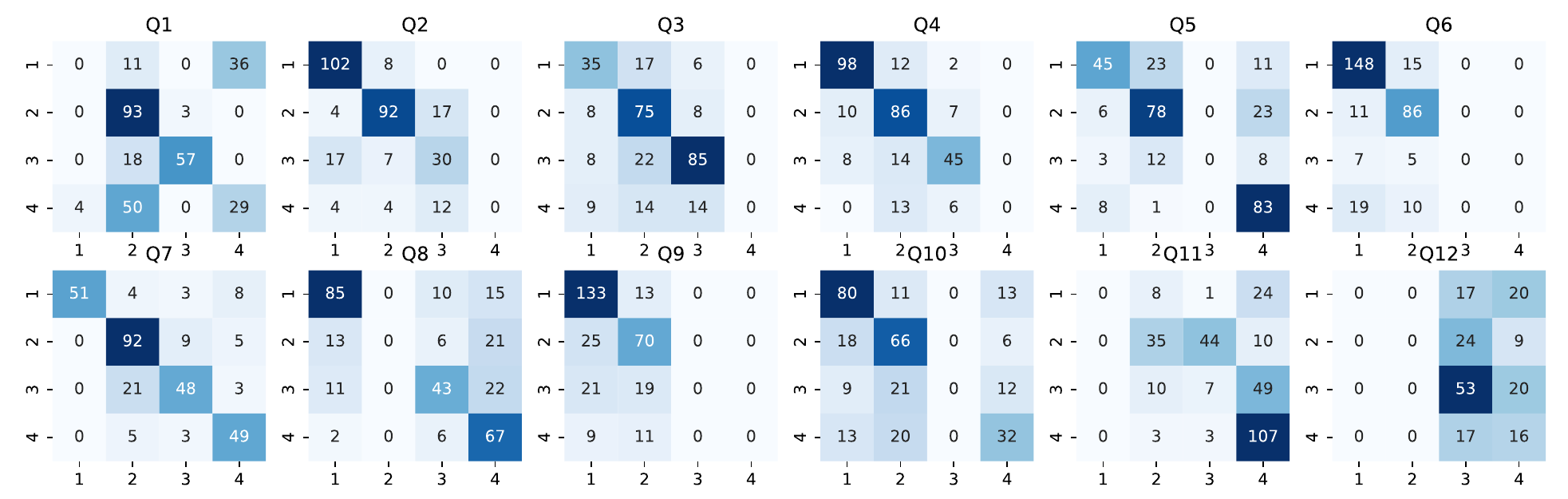}
\caption{Confusion matrix in question answering simulation. Each matrix is one post-course question. X and Y axis represent agents' choices and corresponding real students' choices respectively. There are four choices per question. The confusion matrix shows that question answering simulation has good performance in some questions like Q7 but also bad performance in some questions like Q12.
}
\label{appendix fig: confusion_matrix}
\end{figure*}

Additional experiment results are depicted in Fig. \ref{appendix fig: cdf_gpt_type}, Fig. \ref{appendix fig: cdf_strategy}, Fig. \ref{appendix fig: transcript curve metrics 1}, Fig. \ref{appendix fig: transcript curve metrics 2}, Fig. \ref{appendix fig: truth curve metrics}, Fig. \ref{appendix fig: question ind compare}, Fig. \ref{appendix fig: dur ind compare 0}, Fig. \ref{appendix fig: dur ind compare 1}, Fig. \ref{appendix fig: confusion_matrix}.

\subsection{Additional information of experiment 2}
Here are the details of how we encode all personas and behaviors for evaluation.

For virtual student datasets, we do not have ground truth for comparison. Inspired by existing work \cite{asteriadis2009estimation,brophy1984teacher}
showing that student learning performance is affected by their personalities,
we decide to use Pearson coefficients (similar with \cite{harandi2015effects}) to measure the 
correlation among personas (demographics and student characteristics) and all learning behaviors and outcomes. By doing so, we could measure whether the generated learning behaviors could echo related hypotheses and conclusions of existing student behavioral research to demonstrate the realism of generated behaviors. We examine the following specific aspects: 

\textbf{Persona}: Each characteristic (from learning attitude to family) is either positive (denoted as 1) or negative (denoted as 0). In addition, major and education have several categories, which are normalized to 1. We also encode all characteristics and demographics into one aggregated persona measurement. Specifically, we first normalize each learning characteristic / demographic into the range from 0 to 1. Then we sum all characteristics and demographics, finally divided by the number of all characteristics and demographics, i.e. taking the average of them. The encoded overall persona is a continuous value from 0 to 1.

\textbf{Gaze/Motor}: We use the \textbf{entropy} of the gaze/motor AOI sequences to measure an agents' gaze/motor wandering behaviors. Moreover, we use \textbf{gaze/motor following} to measure whether an agent follows the pace of the lecture closely. For each transcript, gaze/motor following are set to be 1 if agents' gaze/motor AOIs are the same as AOIs of the teacher. Otherwise, they are 0. 
Additionally, we use \textbf{gaze/motor fixing} to measure the extent that agents keep their focus on specific AOIs across transcripts. Gaze/motor fixing are set to be 1 if current gaze/motor AOIs in the current transcript are the same as those in the previous one transcript. Otherwise, they are 0. 
We first calculate these measurements per transcript and then get average results of all transcripts per simulation step.

\textbf{Cognitive States}: We first get cognitive states (workload, curiosity, valid focus, course following, engagement, confusion) generated by agents per transcript and then get average results of all transcripts in one simulation step (slide).

\textbf{Question Answering}: By comparing agents' answers and correct answers, we calculate the average accuracy of all questions in specific simulation step (slide).

\begin{figure*}
\centering
\includegraphics[width=1\linewidth]{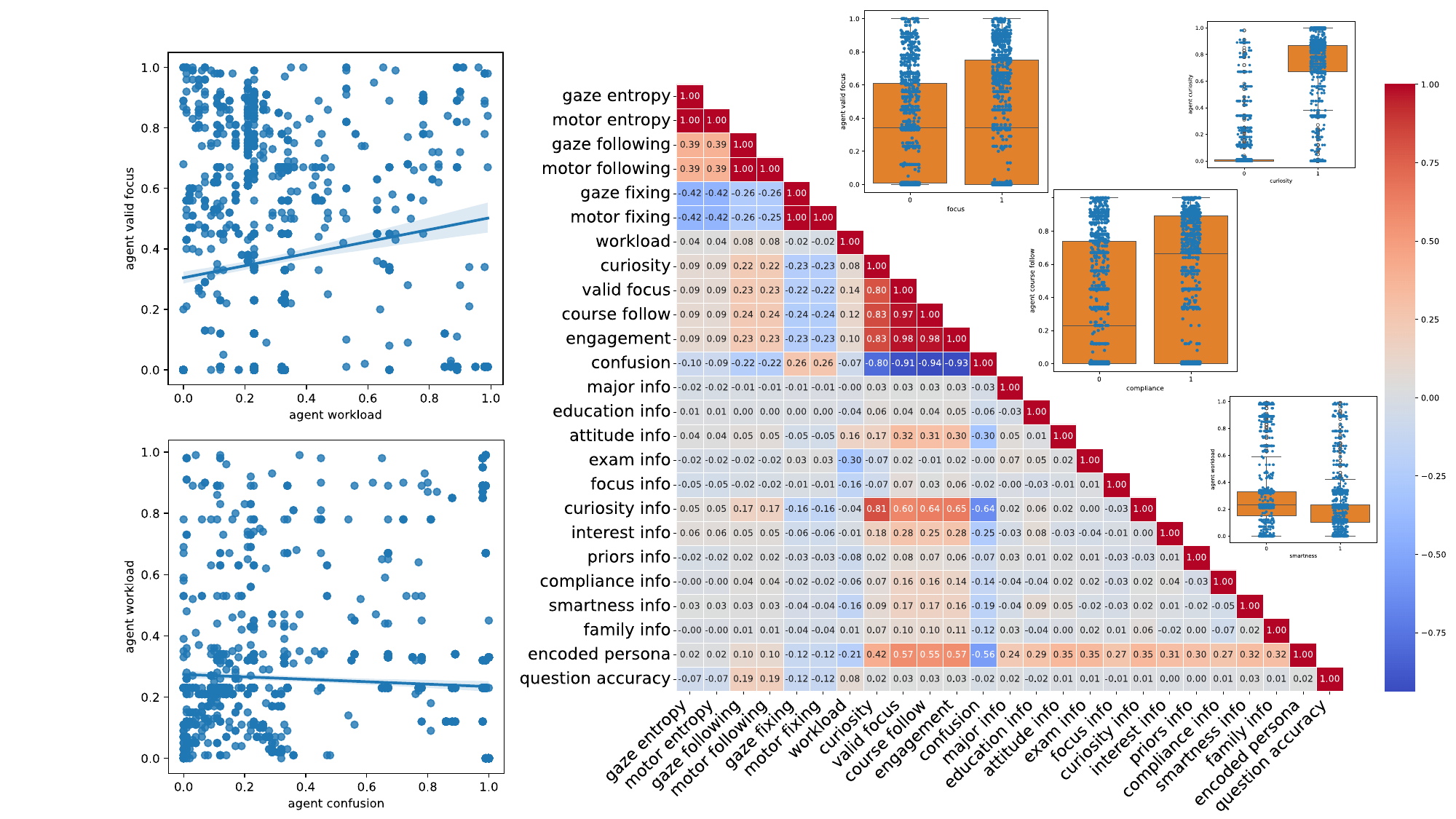}
\caption{Heatmap of correlation matrix as well as examples of correlations of the second experiment using Gemini.
}
\label{appendix fig: heatmap gemini}
\end{figure*}

\begin{figure*}
\centering
\includegraphics[width=1\linewidth]{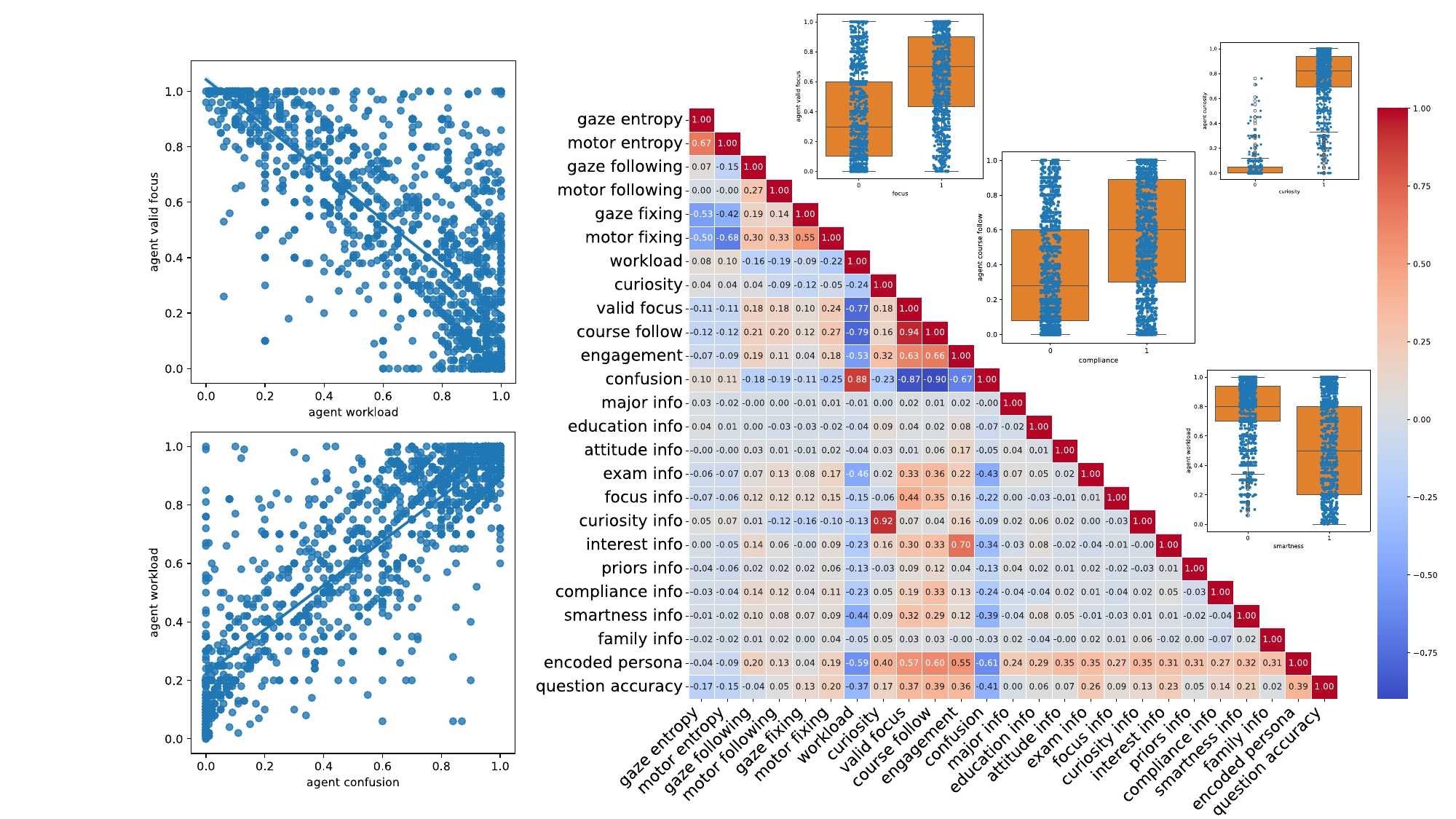}
\caption{Heatmap of correlation matrix as well as examples of correlations of the second experiment using GPT 4.
}
\label{appendix fig: heatmap gpt4}
\end{figure*}

\textbf{GPT4 v.s. Gemini}: We also compare the correlation matrix generated by GPT4-powered student agents (Fig. \ref{appendix fig: heatmap gpt4}) and Gemini-powered student agents (Fig. \ref{appendix fig: heatmap gemini}). As depicted in the two figures, GPT4 could achieve more realistic student behaviors than Gemini.

\begin{table*}[t]
\caption{Configurations of demographics and characteristics of virtual students}
\label{appendix table:demo virtual student}
\vskip 0.15in
\begin{center}
\begin{small}
\begin{sc}
\begin{tabular}{p{3cm}p{10cm}}
\toprule
Category & Items \\
\midrule
Age    &  0: 18-24, 1: 25-31, 2: 32-38, 3: $>$ 39 \\
Gender & 0: female, 1: male, 2: others \\
Major    & 0: Humanities, 1: Social, 2: Natural, 3: Technology, 4: Business, 5: Health\\
Education Level    & 0: high school, 1: undergraduate, 2: master, 3: doctor \\
Learning Attitude    & 1: Very motivated, 0: Not motivated \\
Exam Performance & 1: High GPA, answer test questions correctly, 0: Low GPA. make mistakes in post-test \\
Focus    & 1: Very focus, 0: Usually absent-minded\\
Curiosity    & 1: Curious to explore everything in the course, 0: Not curious at all \\
Interest in Course    & 1: Super interested, 0: Not Interested at all\\
Prior knowledge    & 1: Strong background with prior knowledge, 0: No background without priors \\
Compliance    & 1: Well-behaved to follow teachers, 0: Unwilling to follow teachers \\
Smartness    & 1: Smart to understand everything fast, 0: Not smart, understand things slowly \\
Family    & 1: Parents have a strong academic background, 0: Parents do not care about education \\
\bottomrule
\end{tabular}
\end{sc}
\end{small}
\end{center}
\vskip -0.1in
\end{table*}

\begin{figure*}
\centering
\includegraphics[width=1\linewidth]{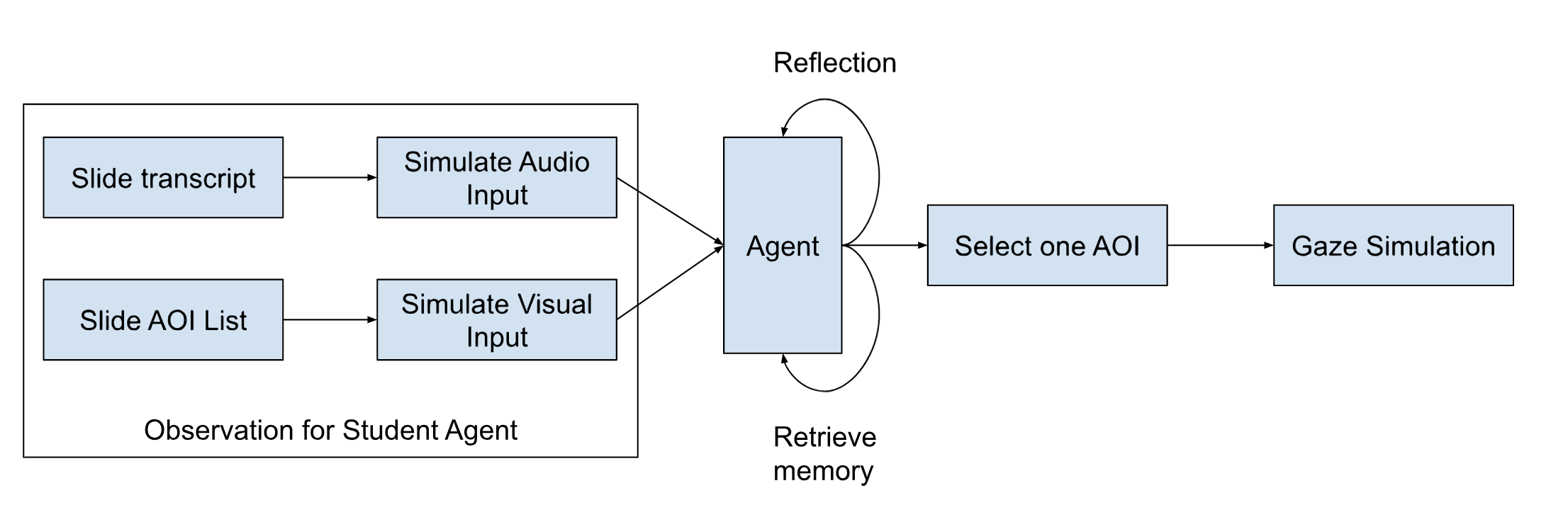}
\caption{Illustration of our way to simulation gaze actions using AOIs.
}
\label{appendix fig: gaze simulation method}
\end{figure*}

\end{document}